\documentclass[twocolumn]{aastex631} 
\usepackage{csvsimple}
\usepackage{multirow,bigdelim}

\usepackage{amsmath}
\usepackage{mathtools}
\usepackage{hyperref}

\newcommand{\T}{{\rm T}}
\newcommand{\msun}{\mbox{$\,{\rm M}_\odot$}}

\usepackage{colortbl}
\definecolor{tablegray}{rgb}{0.89, 0.89, 0.89}
\shorttitle{Exoplanet Aeronomy: WASP-69}
\shortauthors{Levine et al.}

\begin{document}

\title{Exoplanet Aeronomy: A Case Study of WASP-69b's Variable Thermosphere}
\correspondingauthor{W. Garrett Levine}

\email{garrett.levine@yale.edu}

\author[0000-0002-1422-4430]{W. Garrett Levine}
\altaffiliation{DoD NDSEG Fellow}
\affil{Dept. of Astronomy, Yale University. New Haven, CT 06511, USA}

\author[0000-0003-2527-1475]{Shreyas Vissapragada}
\altaffiliation{51 Pegasi b Fellow}
\affil{Center for Astrophysics, Harvard \& Smithsonian, 60 Garden Street, Cambridge, MA 02138, USA}

\author[0000-0002-9464-8101]{Adina~D.~Feinstein}
\altaffiliation{NHFP Sagan Fellow}
\affil{Laboratory for Atmospheric and Space Physics, University of Colorado Boulder, UCB 600, Boulder, CO 80309}

\author[0000-0002-3641-6636]{George W. King}
\affil{Department of Astronomy, University of Michigan, Ann Arbor, MI 48109, USA}

\author[0009-0008-3939-3194]{Aleck Hernandez}
\affil{Department of Physics and Astronomy, Wayne State University, Detroit, MI 48201, USA}

\author[0000-0002-5466-3817]{L\'{i}a Corrales}
\affil{Department of Astronomy, University of Michigan, Ann Arbor, MI 48109, USA}

\author[0000-0002-0371-1647]{Michael Greklek-McKeon}
\affil{Division of Geological and Planetary Sciences, California Institute of Technology, Pasadena, CA, 91125}

\author[0000-0002-5375-4725]{Heather A. Knutson}
\affil{Division of Geological and Planetary Sciences, California Institute of Technology, Pasadena, CA, 91125}

\begin{abstract}
Aeronomy, the study of Earth's upper atmosphere and its interaction with the local space environment, has long traced changes in the thermospheres of Earth and other solar system planets to solar variability in the X-ray and extreme ultraviolet (collectively, ``XUV") bands. Extending comparative aeronomy to the short-period extrasolar planets may illuminate whether stellar XUV irradiation powers atmospheric outflows that change planetary radii on astronomical timescales. In recent years, near-infrared transit spectroscopy of metastable HeI has been a prolific tracer of high-altitude planetary gas. We present a case study of exoplanet aeronomy using metastable HeI transit observations from Palomar/WIRC and follow-up high-energy data from the \textit{Neil Gehrels Swift Observatory} that were taken within one month of the WASP-69 system, a K-type main sequence star with a well-studied hot Jupiter companion. Supplemented by archival data, we find that WASP-69's X-ray flux in 2023 was less than 50\% of what was recorded in 2016 and that the metastable HeI absorption from WASP-69b was lower in 2023 versus past epochs from 2017-2019. Via atmospheric modeling, we show that this time-variable metastable HeI signal is in the expected direction given the observed change in stellar XUV, possibly stemming from WASP-69's magnetic activity cycle. Our results underscore the ability of multi-epoch, multi-wavelength observations to paint a cohesive picture of the interaction between an exoplanet's atmosphere and its host star.\\%
\end{abstract}


\section{Introduction} \label{sec:intro}

Photoevaporative outflows driven by stellar X-ray and extreme UV radiation -- collectively known as ``XUV radiation," with wavelengths $\lambda < 911$\,\AA{} -- may sculpt the observed distribution of exoplanetary radii \citep{owen2019escapeReview}. Indeed, the hot Neptune desert \citep{howard2010occurrence, mazeh2016neptuneDesert} and the radius valley that separates super-Earths from sub-Neptunes \citep{fulton2017gap} are plausibly explained by the XUV-driven destruction of gaseous envelopes \citep{owen2017radiusValley, owen2018subJoviandesert}. Planetary outflows can be revealed by transit spectroscopy of features that probe thermospheric gas, including Lyman-$\alpha$ \citep{vidalmajar2003lyalpha, rockliffe2021LyA}, UV metal lines \citep{vidalmajar2004metals, linsky2010HD209458uv}, H$\alpha$ \citep{yan2018HalphaEscape}, and the metastable HeI triplet near 10830\,\AA{} \citep{seager2000transmission, oklopcic2018windowHe}. While both H$\alpha$ and the metastable HeI triplet are observable from ground-based facilities and do not suffer from interstellar extinction, the former has only been detected in among the hottest, most irradiated giant planets \citep{dossantos2021review}. The metastable HeI feature, in contrast, has been detected from a wider variety of planets; this latter probe has emerged as the most widely-used over the past several years \citep[e.g.][]{nortmann2018wasp69He, allart2018He, salz2018HD189733He, allart2019WASP107, dossantos2020WASP127, kasper2020subNeptuneHe, spake2021WASP107, vissapragada2022upper, kirk2022giantHe,allart2023spirouHe, bennett2023nondetectionWASP48b, zhang2023miniNeptunes, guilluy2023DREAMHe, gullysantiago2024variableTail, guilluy2024GAPSHeI}.

The metastable HeI signal depends sensitively on the high-energy radiation incident on an exoplanet's upper atmosphere. Not only does XUV power photoevaporative outflows \citep{owen2019escapeReview}, but the relative amount of XUV to mid-UV also sets the level populations and ionization environments from which the metastable HeI feature is produced \citep{oklopcic2019radiationHe}. Therefore, connecting the growing body of metastable HeI data to astrophysical models of exoplanet mass-loss requires XUV measurements of host stars. Extreme ultraviolet (EUV) radiation is undetectable by current facilities, so X-ray fluxes are often used to infer total XUV outputs \citep{sanzforcada2011XUV, chadney2015UXV, nortmann2018wasp69He, king2018xuv}. When X-ray data have been unavailable for systems with planetary metastable HeI measurements, previous studies have used proxy spectra from stars with similar masses and activity indices \citep{mansfield2018HATP11, vissapragada2022parker, bennett2023nondetectionWASP48b}. 

Nonetheless, different stars of the same mass can vary by an order of magnitude in X-ray flux \citep{messina2003xray, jackson2012coronalXrayAge, wright2018rotationActivityMdwarf, johnstone2021activeLives}; exoplanet mass-loss rates derived with proxy stellar spectra from another star are similarly uncertain \citep{behr2023muscles}. Compounding the issue, the X-ray output from a given star is often time-variable \citep{wagner1988xrayCycle21, woods2005solarXUV, chadney2015UXV} and should alter exoplanetary atmospheric conditions accordingly \citep{wang2021numericalWASP69}. Thus, single-epoch snapshots of exoplanet thermospheres likely do not reflect the time-averaged values.

Due to this star-to-star variability and time-variability of individual stars' XUV production, a complete understanding of exoplanet aeronomy requires XUV and metastable HeI observations of the same star-planet system across multiple epochs. Individual stellar XUV and planetary thermospheric data are most complementary when obtained simultaneously. In practice, telescope scheduling constraints and transit ephemerides often prohibit this ideal case. Since spot-driven stellar XUV fluxes evolve on timescales a few stellar rotation periods \citep{woods2002solarUVvariability}, we define ``contemporaneous" multi-wavelength observations that occur within such a window as being particularly helpful for the aeronomy science case.

Recent research has pushed towards this multi-wavelength framework, such as a study of HD 189733 by \cite{zhang2022hd189733}. That work found 30\% changes in metastable HeI absorption by the companion hot Jupiter over years-long timescales, which could be explained by the general level of the host's X-ray variability \citep{pillitteri2022xrayHD189733}. However, the high-energy flux timeseries was obtained before any metastable HeI observations. In other multi-wavelength work, \cite{nortmann2018wasp69He}, \cite{czesla2022HATP32}, and \cite{zhang2023miniNeptunes} collected metastable HeI transmission spectroscopy and X-ray fluxes for various systems with varying degrees of contemporaneity.

\begin{table*}[]
\caption{Summary of Palomar/WIRC observations of WASP-69b. ``Start" and ``Finish" correspond to the beginning and end of the science image sequence, respectively. ``Moon Frac." and ``Moon Sep." refer to the illuminated fraction of the Moon and angular on-sky separation between the moon and WASP-69b, respectively. The transit midpoint uncertainty $\sigma_{\text{mid}}$ is calculated from the ephemeris from \cite{kokori2022exoClock}, while $z_{\text{st}}$, $z_{\text{min}}$, and $z_{\text{end}}$ represent the airmasses at the beginning, the minimum, and at the end of the science imaging, respectively. The symbols $n_{\text{exp}}$, $n_{\text{tr}}$, $t_{\text{exp}}$ denote the number of science images, the number of in-transit images, and the exposure time of the science images, respectively.}
\label{tab:palomar}
\centering
\begin{tabular}{|c|c|c|c|c|c|c|c|c|c|c|c|c|}
\hline
\textbf{Night} & \textbf{UT Date} & \textbf{Start} & \textbf{Finish} & \textbf{Moon Frac.} & \textbf{Moon Sep.} & $\sigma_{\text{mid}}$ & \textbf{z$_{\text{st}}$} &  \textbf{z$_{\text{min}}$} & \textbf{z$_{\text{end}}$} & \textbf{$n_{\text{exp}}$} & \textbf{$n_{\text{tr}}$} & \textbf{$t_{\text{exp}}$} \\ \hline
1 & 2019 August 16 & 04:26:06 & 11:01:00 & 99.9\% & 22$^{\circ}$ & 34s & 1.73 & 1.28 & 2.52 & 345 & 117 & 60\,s \\ \hline
2 & 2023 August 25 & 03:26:15 & 08:22:16 & 0.5\% & 67$^{\circ}$ & 54s & 1.94 & 1.28 & 1.44 & 262 & 118 & 60\,s \\ \hline
\end{tabular}
\end{table*}

Here, we present new metastable HeI results from a transit of WASP-69b viewed by the Wide Field InfraRed Camera \citep[WIRC;][]{wilson2003wirc} at Palomar Observatory and high-energy observations from the \textit{Neil Gehrels Swift Observatory} (\textit{Swift}) that were taken within a one-month window. Combining these measurements with archival metastable HeI \citep{nortmann2018wasp69He, vissapragada2020He, vissapragada2022upper, allart2023spirouHe, tyler2023wasp69, guilluy2024GAPSHeI} and high-energy \citep{nortmann2018wasp69He, spinelli2023xray} observations provides a powerful assessment of WASP-69b's outflow variability. The bright ($J$ = 8.0\,mag) host star and large scale height of the hot Jupiter companion \citep[0.26\,M$_{\text{J}}$, 1.06\,R$_{\text{J}}$;][]{anderson2014wasp69} make the WASP-69 system an exceptional target for studies seeking to quantify the time-variability of atmospheric outflows.

Metastable HeI variability was first reported for WASP-69b by \cite{nortmann2018wasp69He}, who uncovered changes in excess absorption within the same month. Then, \cite{allart2023spirouHe} analyzed archival SPIRou data from 2019 and found a lower metastable HeI signal compared to either CARMENES dataset. Later, \cite{tyler2023wasp69} documented changes in post-egress absorption by comparing their Keck/NIRSPEC spectra to CARMENES ones from \cite{nortmann2018wasp69He}. Most recently, \cite{guilluy2024GAPSHeI} presented data from four transits with metastable HeI absorption and simultaneous optical spectroscopy. That study reported and analyzed metastable HeI variability in the context of simultaneous H$\alpha$ measurements. We also take a multi-wavelength approach in our study, but we instead complement our metastable HeI data with observations of stellar XUV flux to connect our research with the geoscientific discipline of aeronomy.

In Section \ref{sec:transit}, we describe our new metastable HeI transit for WASP-69b. Section \ref{sec:model} presents individual light curve fits for both new and archival WIRC observations, a joint fit of WIRC and \textit{TESS} data, and a joint fit of WIRC, \textit{TESS}, and archival spectroscopic data. Then, we analyze the new \textit{Swift} observations of WASP-69 in Section \ref{sec:xray}. With multi-wavelength data, we model the stellar XUV spectrum and hot Jupiter's mass-loss rate via a 1D isothermal Parker wind formulation in Section \ref{sec:variability}. Finally, Section \ref{sec:discussion} discusses the interpretation of variability in the WASP-69 system and elucidates pathways for future research on exoplanet aeronomy.

\section{Metastable He I Transit Observations} \label{sec:transit}

\subsection{Palomar/WIRC Data Collection}

We observed a full transit of WASP-69b with WIRC on the night of 2023 August 25 UT. A summary of the observations for the new light curve and for an archival WIRC light curve from \cite{vissapragada2020He} are provided in Table \ref{tab:palomar}. Conditions were excellent on both observing nights; the humidity was 41\% and 9\% during the new and archival observations, respectively. Before collecting the new light curve, we obtained dark frames, flat-fielding, and HeI arc lamp reference images. Then, we constructed a background frame with a four-point dither immediately prior to taking science data. Our observing setup was identical to that of \cite{vissapragada2020He}, permitting a precise comparison between new and archival transit observations. Science images were obtained with the beam-shaping diffuser (fore wheel) to reshape stellar point spread functions (PSFs) into 3$\farcs$ full-width at half-maximum (FWHM) top-hats \citep{stefansson2017diffuser} and the ultra-narrowband HeI filter (aft wheel).

We positioned the centroid of WASP-69's PSF within 7 pixels ($1\farcs75$) of where the star was placed for the archival WIRC observations. This consistency is particularly important because HeI filter observations have position-dependent wavelength sensitivity \citep{vissapragada2020He}. Imaging was done with an exposure time of 60s, the same integration cadence as Night 1. Guiding was stable throughout the observations. We captured the entire transit as well as approximately 40 and 120 minutes of pre-ingress and post-egress baseline, respectively. For comparison, the archival WIRC observations contain nearly equal pre-ingress and post-ingress baselines (about 120 minutes for each).

\subsection{Light Curve Data Reduction}

We reduced data from the archival (Night 1) and new (Night 2) observations with the pipeline implemented by \cite{vissapragada2022upper}. Raw science images were dark-corrected, flat-fielded, and cleaned of detector cosmetic imperfections. Then, we corrected for telluric emission lines by exploiting the fact that the ultra-narrowband filter has position-dependent absorption. Following the procedure from \cite{vissapragada2020He}, we removed sources in the dither sequence images via sigma clipping and median-scaled the combined dither image for each science frame in 10 pixel radial steps from the point on the filter where the incident light rays are normal to the filter surface. In the light curve modeling, we used the resultant scaling factors from this procedure to detrend on the time-varying telluric water absorption \citep{paragas2021HeHATP18}.

\begin{figure}
    \centering
    \includegraphics[width=1.05\linewidth]{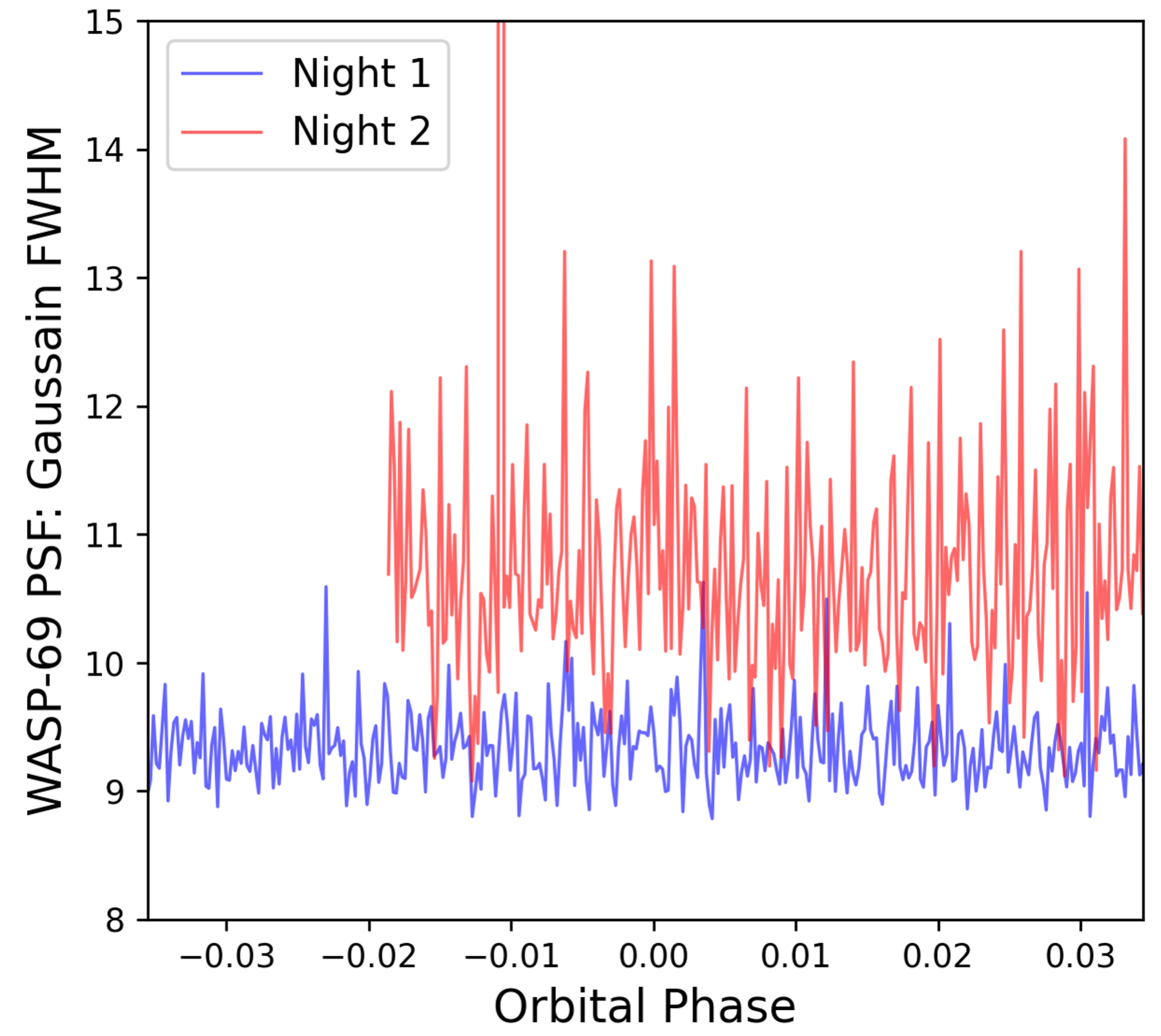}
    \caption{Gaussian FWHM of WASP-69's PSF plotted versus orbital phase during the WIRC science imaging sequence for Night 1 (blue line) and Night 2 (red line). Normalized radial seeing profiles were constructed and fitted with a Gaussian using \texttt{photutils} \citep{bradley2023photutils}. Although the Night 2 PSF varied more than the Night 1 PSF, both timeseries were stable over the observations.}
    \label{fig:radialSeeing}
\end{figure}

Next, we extracted WASP-69's light curve with aperture photometry. First, we selected comparison stars on which to perform differential photometry. Four and two comparison stars in the field were detected beyond 50$\sigma$ significance on Nights 1 and 2, respectively. Previously published Night 1 reductions by \cite{vissapragada2020He} and \cite{vissapragada2022upper} used four comparison stars that were detected beyond 50$\sigma$ significance. Even though the Night 2 data had the same exposure time, only two of those comparison stars met that threshold on Night 2; we attribute this result to minor differences in seeing between the nights (Figure \ref{fig:radialSeeing}) that led to a slightly larger stellar PSF size on Night 2. For all analyses in this paper, we elected to keep only the two comparison stars that were detected beyond 50$\sigma$ significance on both nights. We found, however, that selecting two versus four comparison stars on Night 1 did not meaningfully change the derived transit model in Section \ref{sec:model}.

We determined the optimal aperture size by testing apertures with radii ranging between $3-20$ pixels ($0\farcs75-5\farcs00$) in radius. For each candidate size, we constructed WASP-69's light curve and detrended by the mean of the comparison star fluxes. Local background subtraction around the stars was done with an annulus of $25-50$ pixels ($6\farcs25-12\farcs50$). We then chose the aperture that minimized the root-mean-square (rms) scatter after clipping 5$\sigma$ outliers with a moving median filter. We found optimal aperture sizes of 11 and 12 pixels for Nights 1 and 2, respectively. To ensure that the stellar PSF was stable and consistent during the imaging, we examined the Gaussian FWHM of WASP-69's radial seeing profile for each science exposure. Specifically, we constructed seeing profiles for each image from 100 uniformly-spaced radii between 2-40 pixels and calculated the FWHM with \texttt{photutils}. Figure \ref{fig:radialSeeing} plots the FWHM of WASP-69's PSF versus WASP-69b's orbital phase, assuming the ephemeris published by \cite{kokori2022exoClock}.

Both PSFs were stable across their respective light curves, likely due to the excellent and consistent weather conditions within each night. Although the Night 2 PSF was slightly larger on average than the Night 1 PSF, the FWHM of the Night 2 PSF still remained within 1 pixel of the median value for most exposures. The smaller average FWHM of the Night 1 PSF matches the fact that we found a smaller best-fit aperture (11 pixels) for this timeseries than for the Night 2 PSF (12 pixels).

\section{Multi-Epoch Transit Fitting} \label{sec:model}

\subsection{Palomar/WIRC: Individual Nights} \label{subsec:individual}

With these reduced light curves, we first fit each WIRC dataset independently. Model setups for each night were identical, yielding a direct comparison between the best-fit transit depths. Our procedure was similar to the one described by \cite{vissapragada2022upper}, which uses \texttt{exoplanet} to fit a \texttt{starry}-generated light curve to the data. Our procedure simultaneously fit a light curve model with a model of the systematics. Free parameters in the light curve model were the planet-to-star radius ratio $R_{p}/R_\star$, the exoplanet's orbital period $P$, the reference transit midpoint $t_{0}$, the scaled semi-major axis $a/R_{\star}$, the impact parameter $b$, a jitter $j$ to account for the difference between Poisson noise and the realized scatter, and weights of the comparison stars and covariates for linear systematics detrending. We fixed the orbital eccentricity to zero, in agreement with published radial velocity constraints on WASP-69b that were consistent with a circular orbit \citep[$e < 0.11$;][]{bonomo2017planetMass}.

We fixed limb-darkening coefficients to values that were calculated with \texttt{ldtk} for WASP-69 by \cite{vissapragada2022upper}: $[u_{1}, u_{2}] = [0.38, 0.12]$. Using these constant values, we determined an optimal set of covariates on which to detrend in the final fits. We considered airmass, the centroid offsets, and the water absorption proxy introduced by \cite{paragas2021HeHATP18}. For each combination of these covariates, we optimized the light curve model parameters to find the best-fit solution and removed outlier points more than 4$\sigma$ from the best-fit solution. We iterated this optimization and sigma-clipping until no $4\sigma$ outliers remained. Then, we sampled the posterior with the No U-Turn Sampler (NUTS) in \texttt{pymc3} \citep{hoffman2011NUTS} using four chains with 1500 tuning steps and 3000 draws. We set a target acceptance fraction of 0.99, a high value that is necessary to properly sample topologically complex posteriors.

All priors were informed by the ephemeris from \cite{kokori2022exoClock} and planetary data from \cite{anderson2014wasp69}. Except for the following changes, priors for astrophysical parameters were the same as the ones used for the joint fit in Section \ref{subsec:jointTESS} and provided in Table \ref{tab:fitResults}. Instead of the wide priors for $b$, $a/R_{*}$, $P$, and $t_{0}$ that are listed in Table \ref{tab:fitResults}, we used tighter ones based on literature values selected by \cite{vissapragada2022upper}: $b = \mathcal{N}(0.686, 0.023)$, $a/R_{*} = \mathcal{N}(12.00, 0.46)$, $P = \mathcal{N}(3.8681390, 0.0000006)$, and $t_{0}(\text{BJD} - 2450000) = \mathcal{N}(7176.17789, 0.00017)$. Reference star weights and covariate weights were all assigned uniform $\mathcal{U}(-2, 2)$ priors. We used these prior values to check our results versus the analysis of \cite{vissapragada2022upper}; we later widened these priors when doing joint fits as another robustness check.

After finding the best-fit light curve for a given set of covariates, we calculated the Bayesian Information Criterion (BIC) for that model. Once we obtained BICs for all detrending possibilities, we selected the covariate set for our final fit that provided the best BIC. To fairly compare sets of covariates, we kept a constant jitter term in the fits for covariate selection and ensured that the sigma-clipping procedure in the light curve fit yielded timeseries of the same length. We validated our covariate selection by ensuring that neither the Watanabe-Akaike Information Criterion nor the Pareto-Smoothed Importance Sampling Leave-One-Out (PSIS-LOO) statistic preferred different covariates within their \texttt{arviz}-reported \citep{kumar2019arviz} 1$\sigma$ uncertainties.

With our preferred covariates chosen --- water proxy and airmass for Night 1 and airmass only for Night 2 --- we did a final fit with free values of the limb-darkening coefficients and jitter. Although fixing the limb-darkening coefficients could result in smaller uncertainties on the planet-to-star radii ratio, we decided to remove this external model dependence from our light curve fits. We optimized the solution and sampled the posterior with the same sampling hyperparameters as the aforementioned covariate-finding methodology. Visual inspection of the posterior corner plot indicated that our chains had converged, which we statistically validated by finding a Gelman-Rubin diagnostic $\hat{R} < 1.01$ for each sampled parameter.

We found best-fit $(R_{p}/R_\star) = 0.1470_{-0.0025}^{+0.0027}$ for Night 1 and $(R_{p}/R_\star) = 0.1360_{-0.0032}^{+0.0032}$ for Night 2. These depths differ by 2.6$\sigma$. We confirmed that our results for all derived parameters of the WIRC Night 1 light curve were consistent with the result reported by \cite{vissapragada2022upper}. In Appendix \ref{sec:appendixRobustness}, we analyze how each covariate affected the measured transit depth, assess the degree of correlated noise, and check the fit robustness to changes in aperture sizes and out-of-transit baseline. Comparing the night-to-night results is suggestive of astrophysical variability in WASP-69b's metastable HeI signal, but this procedure cannot rule out the possibility that different best-fit transit shapes masqueraded as two different transit depths. Each timeseries was fit independently, albeit with identical procedures, but a more rigorous test would be to fit the data jointly.

\begin{table}[]
\centering
\caption{Summary of input priors for fitting data from individual nights (Section \ref{subsec:individual}) and joint fitting (Section \ref{subsec:jointTESS}, \ref{sub:jointAll}) along with the resultant posteriors from joint fitting WIRC Night 1, WIRC Night 2, and \textit{TESS} Sector 55. Priors on the ephemeris and planetary parameters were informed by \cite{kokori2022exoClock} and \cite{anderson2014wasp69}, respectively. On the planet-to-star radii ratios $(R_{p}/R_{*})$, the subscripts T, (W,1), and (W,2) correspond to the \textit{TESS}, WIRC Night 1, and WIRC Night 2 datasets, respectively. Quadratic limb-darkening coefficients are given by $u_{1}$ and $u_{2}$, with subscripts T and W for \textit{TESS} and WIRC, respectively. Priors for the limb-darkening coefficients were implemented using the methodology by \cite{kipping2013limbdark} via the \texttt{exoplanet} package. The $j_{\text{W},1}$, $j_{\text{W},1}$, and $\epsilon_{\text{T}}$ are the jitter terms for WIRC Night 1, jitter term for WIRC Night 2, and the \textit{TESS} error scaling parameter.}
\label{tab:fitResults}
\begin{tabular}{|c|c|c|}
\hline
\textbf{Parameter} & \textbf{Prior} & \textbf{Posterior} \\ \hline
$P$ & $\mathcal{N}(3.8681390, 1.0)$ & $3.8681376_{-0.0000018}^{+0.0000018}$\, \text{d}\\ \hline
$t_{0} - 2457000$ & 176.17789 + $\mathcal{U}(0, 1)$ & $176.1783_{-0.0012}^{+0.0012}$ \text{JD}\\ \hline
$a/R_{\star}$ & $\mathcal{U}(5.00, 20.00)$ & $12.21_{-0.12}^{+0.13}$\\ \hline
$b$ & $\mathcal{U}(0, 1)$ & $0.665_{-0.014}^{+0.013}$\\ \hline
$R_{\star}$ & $\mathcal{N}(0.813, 0.056)$ & $0.813_{-0.056}^{+0.057}$ \msun\\ \hline
$(R_{p}/R_{\star})_{\mathrm{T}}$ & $\mathcal{U}(0.00, 0.25)$ & $0.1257_{-0.0010}^{+0.0012}$\\ \hline
$(R_{p}/R_{\star})_{\mathrm{W},1}$ & $\mathcal{U}(0.00, 0.25)$ & $0.1462_{-0.0022}^{+0.0022}$\\ \hline
$(R_{p}/R_{\star})_{\mathrm{W},2}$ & $\mathcal{U}(0.00, 0.25)$ & $0.1337_{-0.0031}^{+0.0032}$\\ \hline
$u_{1,\mathrm{T}}$ & \cite{kipping2013limbdark} & $0.54_{-0.32}^{+0.29}$\\ \hline
$u_{2,\mathrm{T}}$ & \cite{kipping2013limbdark} & $0.048_{-0.350}^{+0.406}$\\ \hline
$u_{1,\mathrm{W}}$ & \cite{kipping2013limbdark} & $0.15_{-0.10}^{+0.15}$\\ \hline
$u_{2,\mathrm{W}}$ & \cite{kipping2013limbdark} & $0.70_{-0.24}^{+0.16}$\\ \hline
$j_{\mathrm{W},1}$ & $\mathcal{LN}(10^{-6}, 10^{-2})$ & $0.00240_{-0.00011}^{+0.00011}$\\ \hline
$j_{\mathrm{W},2}$ & $\mathcal{LN}(10^{-6}, 10^{-2})$ & $0.00301_{-0.00015}^{+0.00016}$\\ \hline
$\epsilon_{\mathrm{T}}$ & $\mathcal{U}(0.5, 1.5)$ & $1.0423_{-0.0091}^{+0.0092}$\\ \hline
\end{tabular}
\end{table}

\subsection{\textit{TESS} + WIRC Joint Fitting} \label{subsec:jointTESS}

Towards our goal of determining whether WASP-69b's thermosphere is time-variable, we implemented a simultaneous fit of the two WIRC light curves along with the light curve from the \textit{Transiting Exoplanet Survey Satellite} \citep[\textit{TESS};][]{ricker2015tess}. This space-based observatory was built to execute an all-sky search for transiting exoplanets and observed WASP-69 in Sector 55 (2022 August 05 UT to 2022 September 01 UT). By fitting the WIRC data with \textit{TESS} broadband photometry, we can self-consistently constrain the transit timing and shape. Thus, this procedure interrogates differences in the WIRC transit depths while controlling for a cause that would be unrelated to the planetary thermosphere. Importantly, our goal in using the \textit{TESS} photometry is not to provide a reference by which to compute excess metastable HeI absorption; other bandpasses that are closer to the 10830\,\AA{} feature are more appropriate for that task (see Section \ref{sec:variability}).

We downloaded the \textit{TESS} 20s cadence Presearch Data Conditioning Simple Aperture Photometry \citep[PDCSAP;][]{TESSFastLCs} times, fluxes, and flux uncertainties from the mission's Science Processing Operations Center (SPOC) pipeline \citep{jenkins2016SPOC} via \texttt{lightkurve}, which queries the Mikulski Archive for Space Telescopes (MAST) server. While seven transits were expected to occur within the sector based on the ephemeris published by \cite{kokori2022exoClock}, two of those events occurred during gaps in data coverage.

\begin{figure}
    \centering
    \includegraphics[width=0.9\linewidth]{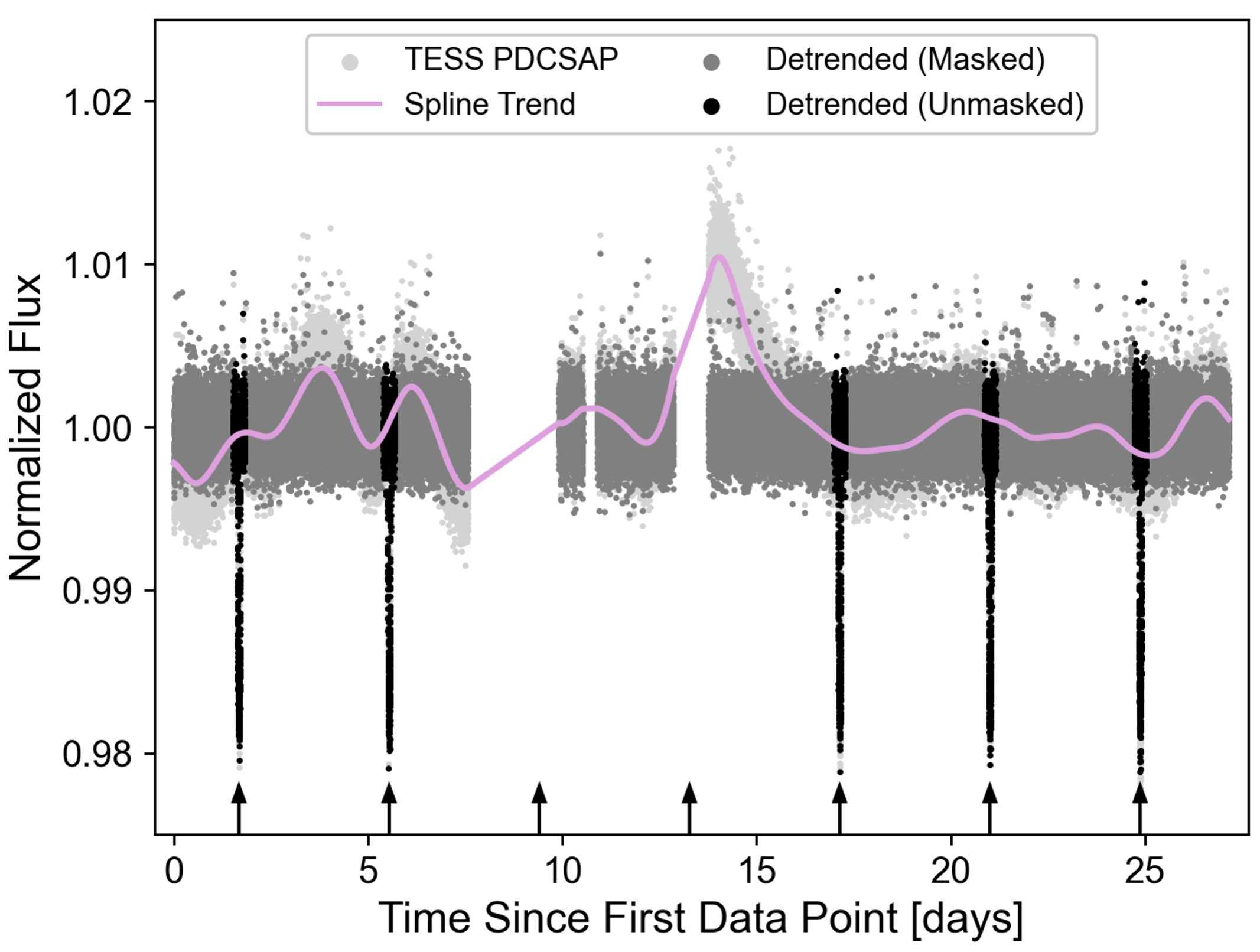}
    \caption{Photometry of WASP-69 from \textit{TESS} Sector 55. The lightest grey points are the PDCSAP flux that were downloaded from MAST, and the magenta line shows the best-fit spline from \texttt{Keplerspline}. Detrended flux, obtained by dividing the PDCSAP flux by the spline, is shown with two colors. Dark grey dots are detrended points far from the observed transits that are removed by the mask. Black dots are detrended points that are included by the mask. Upward-pointing arrows indicate expected transit midpoints from orbital parameters published by \cite{kokori2022exoClock}.}
    \label{fig:tessDetrend}
\end{figure}

The PDCSAP photometry exhibits secular trends which are likely attributed to stellar variability and/or instrumental effects. To detrend photometry while preserving transits, we fit a spline with break point spacing of 5 days to the PDCSAP points with \texttt{Keplerspline} v2 code\footnote{\url{https://github.com/avanderburg/keplersplinev2}} \citep{vanderburg2014K2, shallue2018deepLearningExoplanet}. Dividing the original data by the spline trend flattened the light curve. We illustrate these pre-processing steps in Figure \ref{fig:tessDetrend}. WASP-69b's transits are deeper than any stellar variability or instrumental systematics, so we did not sigma-clip outliers until the joint-fitting itself. To reduce the computational burden of the joint fit, we filtered to keep only points that were within 0.15\,d of any observed transit midpoints.

\begin{figure}
    \centering
    \includegraphics[width=1\linewidth]{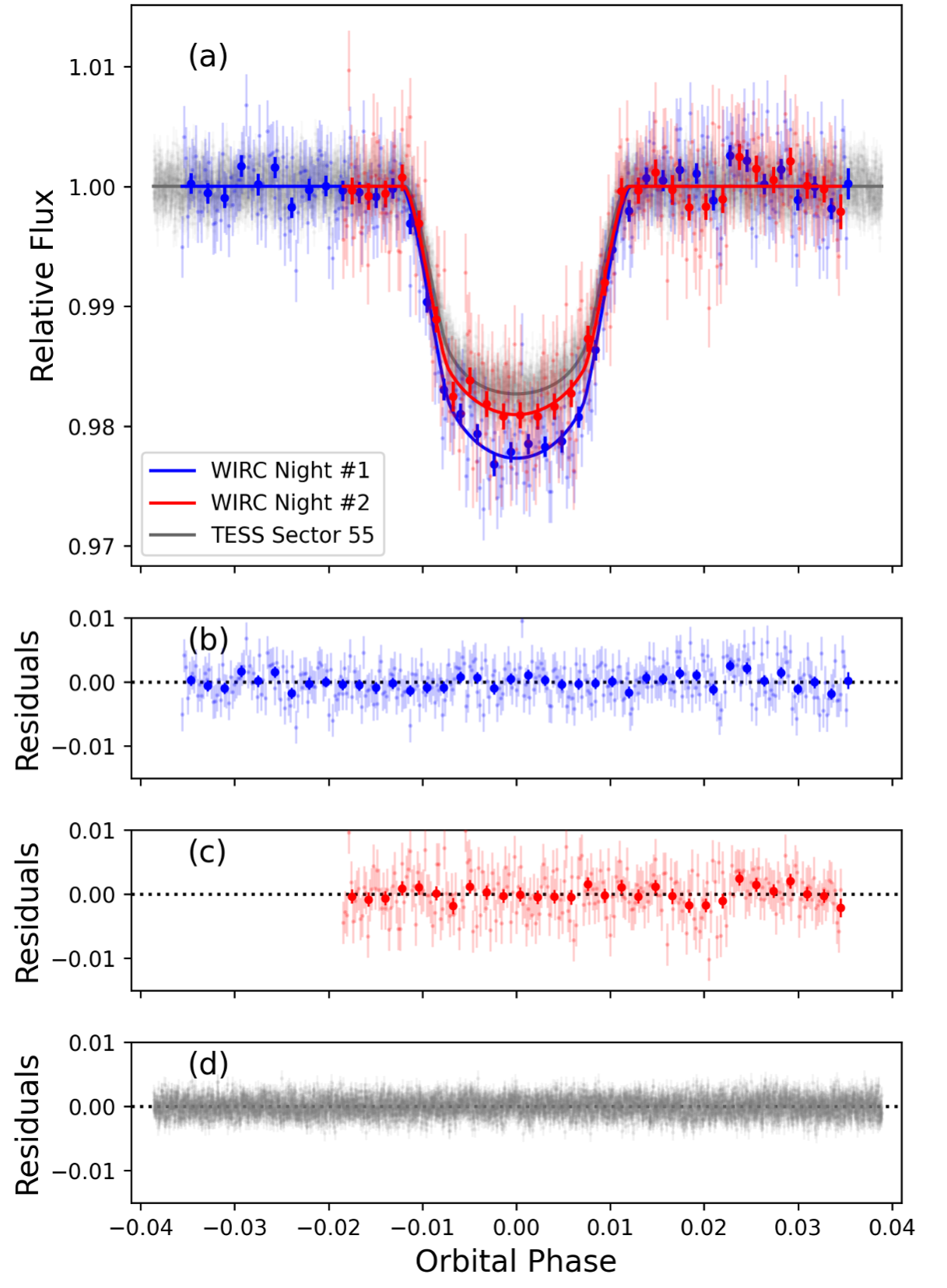}
    \caption{Results from joint-fitting WIRC Night 1 (blue), WIRC Night 2 (red), and \textit{TESS} (grey) data. Panel (a) shows the phase-folded light curves for all three timeseries. Individual points and their uncertainties are shown with small markers, and Palomar/WIRC data binned to ten-minute intervals are shown with larger markers. Best-fit light curve models are shown as solid lines. Panels (b), (c), and (d) show the residuals for the WIRC Night 1, WIRC Night 2, and \textit{TESS} timeseries, respectively.}
    \label{fig:jointFit}
\end{figure}

In our joint model, we required that all three light curves share the same orbital period $P$, transit midpoint reference time $t_{0}$, scaled semi-major axis $a/R_{*}$, impact parameter $b$, and stellar radius $R_{*}$. We allowed different quadratic limb-darkening coefficients between the \textit{TESS} $(u_{1,\text{T}}, u_{2,\text{T}})$ and WIRC $(u_{1,\text{W}}, u_{2,\text{W}})$ bandpasses. To assess the time-variability of the metastable HeI signal, we independently fit for the planet-to-star radii ratio in the \textit{TESS} bandpass $(R_{p}/R_{\star})_{\text{T}}$, the first WIRC night $(R_{p}/R_{\star})_{\text{W},1}$, and the second WIRC night $(R_{p}/R_{\star})_{\text{W},2}$. Each WIRC time series could have different values for jitter, weights for comparison stars, and weights for shared covariates. We also included an error-scaling parameter for the \textit{TESS} photometry to account for deviations from pure photon noise; this term is similar in function to the WIRC jitter terms. Finally, we kept the same optimal sets of covariates for each night as were determined in Section \ref{subsec:individual}.

When comparing ultra-narrowband photometry from multiple epochs, transit depth variability can emerge from the changing topocentric radial velocity of the target star. If the metastable HeI triplet falls on parts of the bandpass with different transmission properties in two different observations, then the recorded transit depths could differ even if the planetary spectrum is unchanged. This issue was negligible between our two WIRC timeseries, as the difference in WASP-69's topocentric radial velocity was less than $5$\,km\,s$^{-1}$ at the respective transit midpoints. The 10830 \AA{} wavelength was shifted in the observatory frame by 0.2 \AA{}, much smaller than the WIRC filter's FWHM of 6.35\,\AA. Nevertheless, we accounted for this shift when we analyzed variability among a larger set of observations in Section~\ref{sub:jointAll}.

\begin{figure}
    \centering
    \includegraphics[width=1\linewidth]{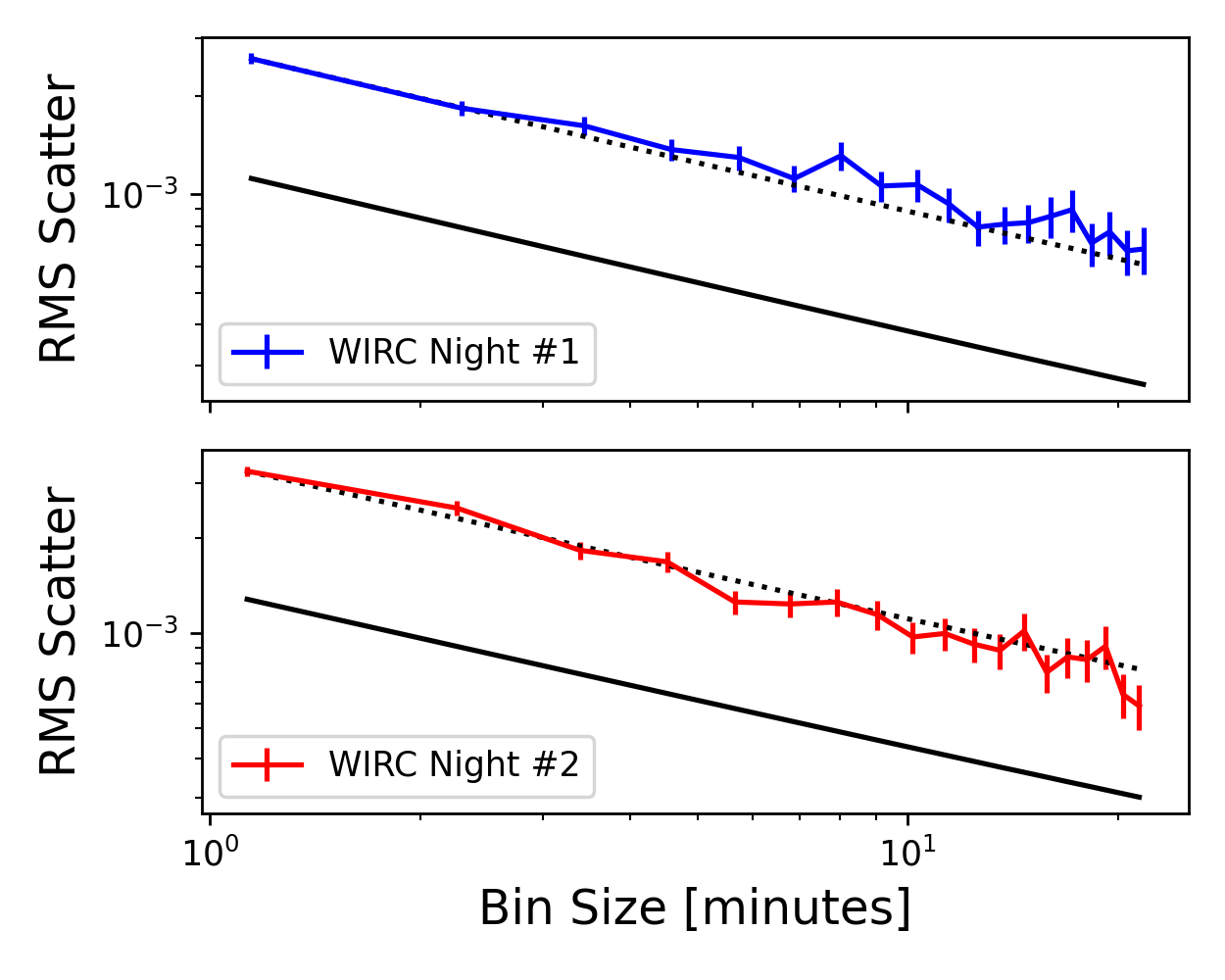}
    \caption{Allan deviation plots for each detrended WIRC light curve in the joint fitting with both time series and \textit{TESS} data. The top and bottom panels show results from the detrended Night 1 and Night 2 light curves, respectively. Colored lines with errorbars denote the rms error of the binned residuals for each light curve. The solid black lines denote the expected noise from purely Poisson statistics, and the dotted line shows the photon noise scaled-up to the rms error of the binned residuals.}
    \label{fig:allanJoint}
\end{figure}

With this model setup, we first removed 10$\sigma$ outliers from the reduced WIRC photometry. Then, we maximized the log-likelihood of the joint model with \texttt{pymc3} and executed sequential rounds of sigma-clipping to eliminate points from both the WIRC and \textit{TESS} timeseries whose residuals were at least 5$\sigma$ discrepant from the maximum \textit{a posteriori} (MAP) solution. When no $5\sigma$ outliers remained, we re-optimized the joint model. Finally, we sampled the posteriors with \texttt{NUTS} to obtain the posterior probability distributions. We ran 4 chains with 1000 tuning steps and 5000 draws. As in Section \ref{subsec:individual}, we set a target acceptance fraction of 0.99 and verified that $\hat{R} < 1.01$ for each parameter in the posteriors.

\begin{figure}
    \centering
    \includegraphics[width=0.8\linewidth]{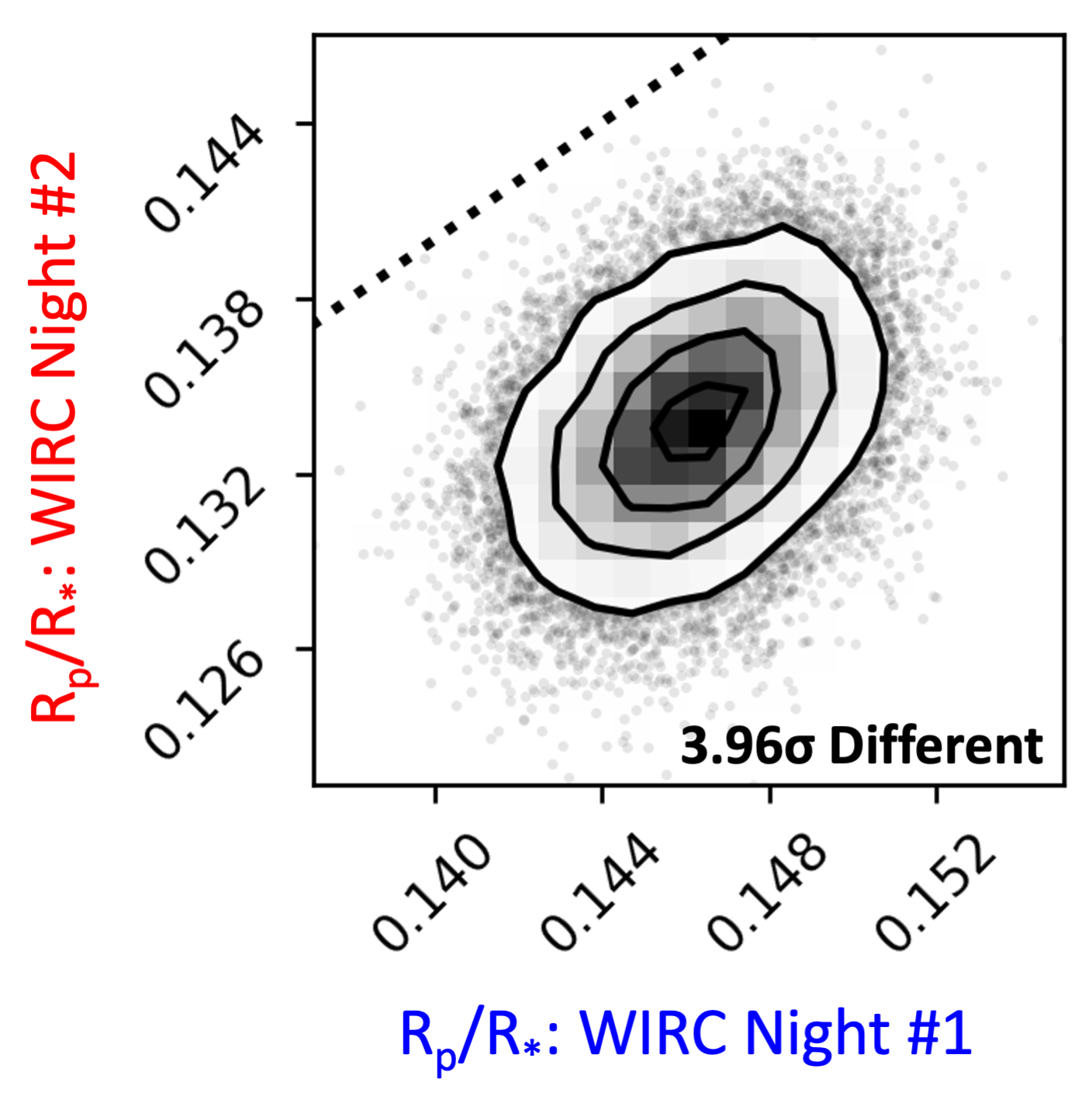}
    \caption{Two-dimensional histogram of the posterior distributions for $(R_{p}/R_{\star})_{\text{W},2}$ versus $(R_{p}/R_{\star})_{\text{W},1}$. The dotted line represents equal planet-to-star ratios for the two metastable HeI transits obtained with WIRC.} 
    \label{fig:cornerRpRs}
\end{figure}

Derived parameters of the joint fit are presented in Table \ref{tab:fitResults} along with best-fit light curves and residual plots in Figure \ref{fig:jointFit}. The Allan deviation plots in Figure \ref{fig:allanJoint} show that the residuals do not contain significant correlated noise. Another indicator of fit robustness is that the posterior distributions of parameters where we used wider priors versus the ones from Section \ref{subsec:individual} -- namely $b$, $a/R_{*}$, $P$, and $t_{0}$ -- converged in these fits to being consistent with those literature-derived tighter priors. Although the WIRC transit depths would be 3.21$\sigma$ discrepant if their posteriors were uncorrelated, inspecting the samples drawn from the distribution (Figure \ref{fig:cornerRpRs}) reveals that the $(R_{p}/R_{\star})_{\text{W}}$ values are not independent. To account for this positive correlation, we computed the covariance matrix between the vectors of posterior samples for the $(R_{p}/R_{*})$ values and calculated the variance while including the off-diagonal terms. With this procedure, we found that the actual difference is 3.96$\sigma$. This result reinforces our conclusion from analyzing each night separately: WASP-69b's metastable HeI absorption signal differed between the two WIRC-observed transits.

The correlation between the WIRC planet-to-star radius ratios is explained by our requirement that the transit shape remain constant between the two transits. Changes in $a/R_{*}$, $b$, or the limb-darkening coefficients that affect one transit depth affect the other one in the same direction. The naive assumption of independent error bars underestimates the true statistical significance of transit depth variability between the WIRC nights, and the joint model reveals more significant variability in the metastable HeI absorption.

\subsection{Assessing Variability With Spectroscopic Data} \label{sub:jointAll}

In addition to the ultra-narrowband photometry from WIRC, WASP-69b's metastable HeI signal has been probed spectroscopically by numerous works. Two transits were observed using CARMENES \citep{nortmann2018wasp69He}, one transit was observed using SPIRou \citep{allart2023spirouHe}, one transit was observed using Keck/NIRSPEC \citep{tyler2023wasp69}, and four transits were observed using GIANO-B \citep{guilluy2024GAPSHeI}. These spectroscopic timeseries\footnote{Timeseries of reduced and published spectra were kindly provided to us by the lead authors of the aforementioned papers.} provide further data through which variability can be probed. We therefore implemented another joint model, this time with these spectroscopic metastable HeI data, the WIRC data, and \textit{TESS} photometry. Since \cite{guilluy2024GAPSHeI} noted that the transit from 2021 October 28 UT was contaminated by telluric OH emission, we excluded that dataset from our analysis; thus, we only examined three GIANO-B transits.

Unlike spectroscopic data with resolution $R \gtrsim 40$k, the WIRC HeI filter has an effective $R \sim 1700$ and can only detect the underlying triplet convolved with the filter's transmission profile \citep{vissapragada2020He}. Thus, comparing spectroscopic data and WIRC light curves requires translating the velocity-resolved observations into simulated WIRC-like photometry \citep{vissapragada2020He}. Alternatively, we could have attempted to translate the WIRC observations into a velocity-resolved profile, but this procedure would have required undue assumptions on the underlying spectral feature.

WASP-69's topocentric radial velocity among the spectroscopic datasets differs by more than 30\,km\,s$^{-1}$, which would Doppler shift the wavelengths on the narrowband filter by more than 1\,\AA{}. To eliminate this non-astrophysical cause of transit depth variability, we Doppler shifted all spectroscopic data as though the observations were being conducted from WIRC on Night 1. Shifting to Night 2 instead does not tangibly affect the final fit since the WIRC radial velocities were close. 

Starting from each of the reduced spectroscopic timeseries, we calculated the excess absorption $f_{\text{ex}}$ for each spectrum in the WIRC bandpass using Eq. 3 from \cite{vissapragada2020He}:

\begin{equation} \label{eq:convolve}
    f_{\text{ex}} = \frac{\int f_{ex,\lambda}T_{\text{W},\lambda}d\lambda}{\int T_{\text{W},\lambda}d\lambda}
\end{equation}

\noindent where $T_{\text{W},\lambda}$ is the WIRC ultranarrowband filter's transmission and $f_{\text{ex}, \lambda}$ is the excess absorption, both at wavelength $\lambda$. Both integrals in Equation \ref{eq:convolve} were computed over the same wavelength range.

Then, we constructed WIRC-like light curves by combining the WIRC-like excess depth timeseries and a reference transit depth of $(R_{p}/R_{*})_{\text{ref}} = 0.1286 \pm 0.0002$ in the 11108-11416\,\AA{} bandpass from the \textit{Hubble Space Telescope's} (\textit{HST}'s) Wide Field Camera 3 presented by \cite{tsiaras2018giantPlanet}. These wavelengths are closer to the 10830\,\AA{} triplet than the \textit{TESS} bandpass, making \textit{HST} a more appropriate reference. This benchmark transit depth was also used for WASP-69b by \cite{vissapragada2020He}. Although planetary absorption from water is present in this \textit{HST} bandpass, this spectral feature is small in magnitude versus our transit depth precision and does not affect our analysis \citep{vissapragada2022upper}.

We estimated uncertainties on the WIRC-like light curve points through a Monte Carlo procedure, where we generated 1000 trial excess absorption timeseries for each spectral timeseries. In each trial, we perturbed each pixel-by-pixel normalized flux by a Gaussian centered on zero and with standard deviation of the uncertainty for that pixel. Pixel-level uncertainties were only provided with the CARMENES and SPIRou data in each spectrum, so we estimated the uncertainty in each spectrum from the Keck/NIRSPEC and GIANO-B timeseries by calculating the rms scatter of the pixels from 10820-10830\,\AA{} and 10836-10839.5\,\AA, respectively.

\begin{table*}[]
\centering
\caption{Summary of derived metastable HeI planet-to-star radius ratios and excess depths, where both values are calculated in the WIRC filter's bandpass from joint fitting all metastable HeI data with \textit{TESS} photometry. The reference transit depth was taken from \textit{HST} results reported by \cite{tsiaras2018giantPlanet}. The CARMENES, Keck, WIRC Night 1, SPIRou, and GIANO timeseries were first published by \cite{nortmann2018wasp69He}, \cite{tyler2023wasp69}, \cite{vissapragada2020He}, \cite{allart2023spirouHe}, and \cite{guilluy2024GAPSHeI} respectively. WIRC Night 2 is the new transit presented in this work.}
\label{tab:transitDepthsAll}
\begin{tabular}{|c|c|c|c|}
\hline
\textbf{Timeseries} & \textbf{Date (UT)} & \textbf{WIRC-like $(R_{p}/R_{*})$} & \textbf{Excess Depth [\%]} \\ \hline
CARMENES \#1 & 2017 August 22 & $0.1444_{-0.0018}^{+0.0018}$ & $0.43_{-0.05}^{+0.05}$ \\ \hline
CARMENES \#2 & 2017 September 22 & $0.1392_{-0.0021}^{+0.0020}$ & $0.28_{-0.06}^{+0.06}$ \\ \hline
Keck/NIRSPEC & 2019 July 12 & $0.1396_{-0.0019}^{+0.0019}$ & $0.30_{-0.05}^{+0.05}$ \\ \hline
GIANO Night 1 & 2019 July 24 & $0.1503_{-0.0071}^{+0.0067}$ & $0.61_{-0.21}^{+0.20}$ \\ \hline
WIRC Night 1 & 2019 August 16 & $0.1468_{-0.0023}^{+0.0023}$ & $0.50_{-0.07}^{+0.07}$ \\ \hline
SPIRou & 2019 October 13 & $0.1376_{-0.0070}^{+0.0067}$ & $0.23_{-0.19}^{+0.18}$ \\ \hline
GIANO Night 2 & 2020 August 09 & $0.1410_{-0.0057}^{+0.0055}$ & $0.33_{-0.16}^{+0.15}$ \\ \hline
GIANO Night 3 & 2022 September 14 & $0.1429_{-0.0044}^{+0.0044}$ & $0.38_{-0.13}^{+0.13}$ \\ \hline
WIRC Night 2 & 2023 August 25 & $0.1343_{-0.0032}^{+0.0031}$ & $0.15_{-0.09}^{+0.09}$ \\ \hline
\end{tabular}
\end{table*}

Next, we added these WIRC-like data into another joint fit model based on the procedure from Section \ref{subsec:jointTESS}. We required that all light curves --- \textit{TESS}, WIRC, and WIRC-normalized spectroscopic data --- share global parameters of the orbit: period $P$, impact parameter $b$, and transit midpoint reference time $t_{0}$. All metastable HeI light curves were required to use the same limb-darkening coefficients, but the planet-to-star radius ratio for each metastable HeI transit was allowed to vary independently. Each excess depth timeseries from the spectroscopic observations was given a free jitter parameter but was otherwise not detrended. For computational efficiency, we binned the \textit{TESS} data from the original 20\,s cadence into 200\,s intervals for only this fit; this coarser broadband timeseries reduced the posterior sampling runtime.

To fit the light curve model, we followed the same iterative optimization, sigma-clipping, and posterior sampling procedures as described in Section \ref{subsec:jointTESS} and treated all metastable HeI data as WIRC-like light curves. During this procedure, we noticed that the post-egress tails of escaping gas reported from Keck/NIRSPEC by \cite{tyler2023wasp69} and from GIANO-B by \cite{guilluy2024GAPSHeI} may have been detectable by WIRC. The light curve model that we implemented fits a symmetric transit shape, so including post-midpoint data could have spuriously depressed the best-fit transit depth. To treat all spectroscopic data fairly and identically, we therefore truncated all the spectroscopic timeseries to only consider times before the transit midpoint. By only considering the first part of the spectroscopic transits, we selected the data that is least affected by the post-egress tail. Notably, \cite{vissapragada2020He} demonstrated that the post-egress tail detected by \cite{nortmann2018wasp69He} with CARMENES was not detectable at WIRC's resolution.

\begin{figure*}
    \includegraphics[width=0.95\linewidth]{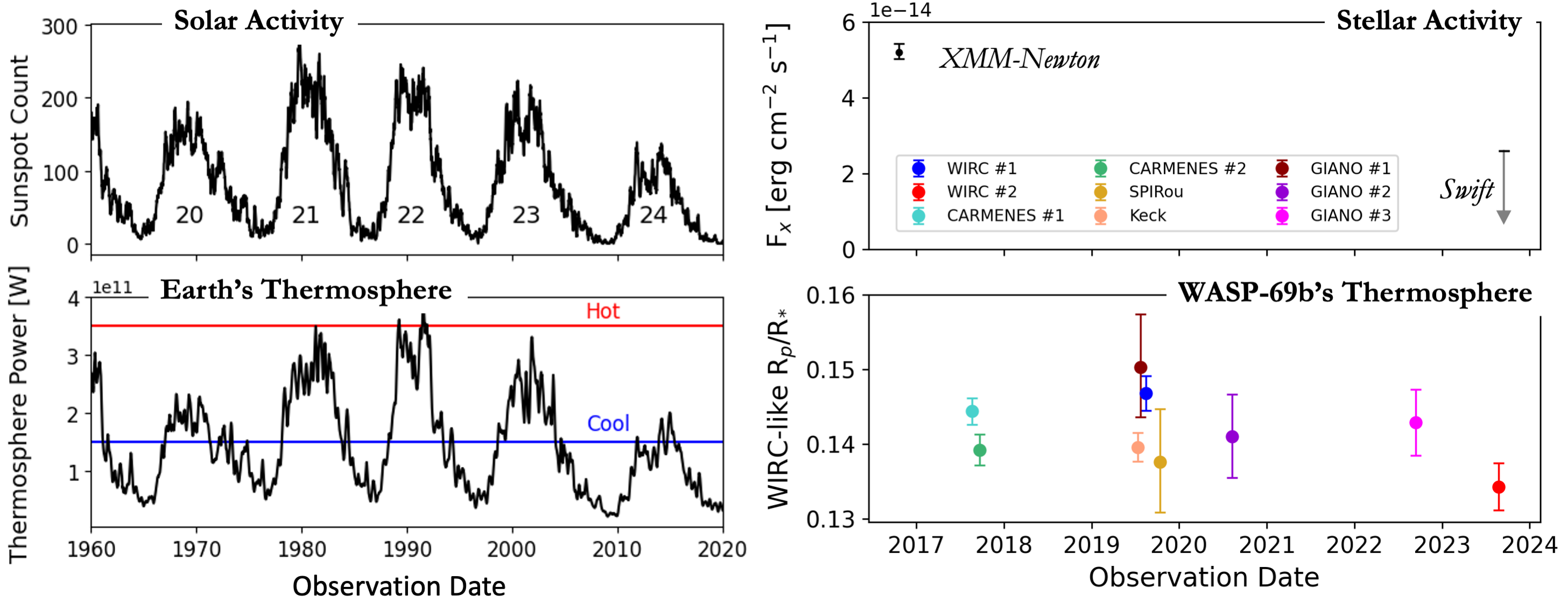}
    \caption{Summary of comparative aeronomy between the terrestrial thermosphere and WASP-69b. Left-Hand Side: Timeseries of solar activity indicator (top panel) and terrestrial thermospheric conditions (bottom panel) from 1960-2020. Numbers on the top panel label the solar cycles by their numbers. Right-Hand Side: Analogous plots for the WASP-69 system. The top panel shows stellar activity proxied by X-ray flux observed at Earth, and the bottom panel shows WASP-69b's thermospheric conditions as traced in metastable HeI from Section \ref{sub:jointAll}. Multi-epoch, multi-wavelength campaigns are necessary to extend the well-developed field of solar system aeronomy to the exoplanets.}
    \label{fig:aeronomy}
\end{figure*}

Finally, we sampled the posterior using \texttt{NUTS} with the same procedure as Section \ref{subsec:jointTESS}. Table \ref{tab:transitDepthsAll} summarizes the results from this joint model, and Figure \ref{fig:aeronomy} (bottom right panel) shows a timeline of the WIRC-normalized transit depths. Light curves and residuals for all metastable HeI timeseries are located in Appendix \ref{sec:appendixCorner}. We found that each of the actual WIRC transit depths were consistent within the $1\sigma$ uncertainties of the transit depth results for those timeseries from the joint fit with only \textit{TESS} data in Section \ref{subsec:jointTESS}. A corner plot with detailed information on the fits is provided in Appendix \ref{sec:appendixCorner}. In the posterior, the two most discrepant metastable HeI transit depths are the two WIRC transit depths ($4.0\sigma$). The first and second CARMENES transit depths are $2.8\sigma$ and $1.3\sigma$ discrepant with the transit depth from WIRC Night 2, respectively. Taken together, this comprehensive light curve model with archival spectroscopic data on the planet demonstrates that its metastable HeI signal is variable over timescales of several years.

\section{\textit{Swift} Observations and Results} \label{sec:xray}

\subsection{X-Ray Telescope Data}


To understand the space weather environment in which the August 2023 metastable HeI signal was produced, we collected high-energy data on WASP-69 with the \textit{Neil Gehrels Swift Observatory} (target ID: 34059). This satellite, previously known as the \textit{Swift Gamma-Ray Burst Explorer}, is a NASA spacecraft that was launched into low-Earth orbit in 2004 with the primary purpose of studying gamma-ray bursts and their afterglows \citep{SwiftMissionGRB}. Although exoplanet astronomy was only emerging during the mission development, \textit{Swift's} ability to measure high-energy radiation has since contributed to the understanding of host stars \citep{bourrier2020HD189733, becker2020TRAPPIST1, spinelli2023UVHabitable} and the escape of short-period exoplanet atmospheres \citep{salz2019swift, corrales2021Swift}.

We collected \textit{Swift} data on WASP-69 between 2023 September 13 UT and 2023 September 19 UT as a Target of Opportunity (ToO; as outlined in Table \ref{tab:swift}) with 11.9ks of total integration time. This timeframe is contemporaneous with the WIRC Night 2 observations of metastable HeI by our definition from Section \ref{sec:intro} since WASP-69b's rotation period is approximately 23\,d \citep{bonomo2017planetMass}. All observations consisted of simultaneous observations with the X-Ray Telescope \citep[XRT;][]{SwiftXRT} and monitoring with the Ultra-Violet Optical Telescope \citep[UVOT;][]{SwiftUVOT} in the UVM2 filter. \textit{Swift} had previously visited WASP-69 once in 2015 for approximately 2ks (observation ID: 00034059001). That short integration cannot provide meaningful constraints in the context of our scientific objective, so we did not analyze those archival data.

Upon creating a stacked XRT image of all five \textit{Swift} visits from 2023, we did not see any readily apparent sources by-eye. There was a small cluster of counts within a 20" radius near the location of WASP-69, indicating a possible marginal detection. To check our intuition, we ran a source detection algorithm from \texttt{Ximage} in \texttt{HEASoft} (version 6.28). We found that WASP-69 was detected at $2\sigma$ confidence but not at the 3$\sigma$ level, reinforcing the notion of a marginal detection.

Given the marginal detection, the most robust constraint on WASP-69's X-ray flux requires an upper limit analysis. Using the \texttt{astropy.stats} module, we calculated a Poissonian confidence interval based on the Bayesian approach of \citet{Kraft1991}. This yielded a 95\% confidence interval on the number of source counts to be [0--11.5], with the upper end of this interval corresponding to a count rate of 0.9\,ks$^{-1}$. To convert this result to a X-ray flux from WASP-69, we scaled the two-temperature APEC model fit to the 2016 \textit{XMM-Newton} observation by \citet{spinelli2023xray} so that the model \textit{Swift} count rate indicated by \texttt{Xspec} matched our statistical upper limit for the measured count rate. While X-ray spectral shapes can change with varying flux, especially during flares, we did not detect any flares (see Section \ref{sub:uvot}). We found the 95\% upper limit on the 0.3--2.4\,keV unabsorbed flux to be 2.6$\times 10^{-14}$\,erg\,cm$^{-2}$\,s$^{-1}$. For comparison, the best-fit model published by \citet{spinelli2023xray} from \textit{XMM-Newton} observations in 2016 predicts a flux of $5.19^{+0.23}_{-0.17} \times 10^{-14}$\,erg\,cm$^{-2}$\,s$^{-1}$ in the same 0.3--2.4\,keV energy range. The archival \textit{XMM-Newton} flux is two times larger than the \textit{Swift} XRT upper limit from 2023, and we show these data on the timeline in Figure \ref{fig:aeronomy}.

\subsection{Ultraviolet Optical Telescope Results} \label{sub:uvot}

We examined the \textit{Swift}/UVOT light curve of WASP-69 for flares. All data were obtained with the UVM2 filter, which covers the 200-250~nm wavelength range. This bandpass is particularly important because the ionization energy of metastable HeI $2^3S_1$ is 4.767~eV, making wavelengths close to and shortward of 260~nm efficient at removing metastable HeI through photoionization. Since 2023~August, a malfunction in one of the \textit{Swift} inertial reference units has degraded its attitude control, introducing a source of instrumental jitter. A visual inspection of the UVOT images demonstrated that the pointing instability did not exceed a few arcseconds. We followed standard UVOT data reduction procedures to stack images and ran \texttt{uvotsource} to do aperture photometry. To account for the increased apparent size of the UVOT image point sources caused by the degraded attitude control, we increased the photometric aperture region to have a $15''$ radius with a $20-35''$ background annulus. 

We found that WASP-69 has an apparent magnitude of $18.16 \pm 0.02$ in the UVM2 band, equivalent to an apparent flux of $1.20 (\pm 0.02) \times 10^{-15}$~erg~s$^{-1}$~cm$^{-2}$~A$^{-1}$. 
We used \texttt{uvotmaghist} to extract an NUV light curve for each UVOT observation, using the same photometric aperture described above. No significant NUV flaring activity was detected. The consolidated light curve hints at a slight downward trend in the NUV flux over the course of our observing campaign, but the change in flux is around $5\%$, similar in magnitude to the error estimates for individual flux values in the \texttt{uvotmaghist} light cruve. Thus, the observed light curve is relatively consistent with a constant NUV flux.

\begin{table}[]
\centering
\caption{Summary of Swift ToO observations of WASP-69. All UT dates and times correspond to September 2023.}
\label{tab:swift}
\begin{tabular}{|c|c|c|c|}
\hline
\textbf{Obs ID} & \textbf{UT Start} & \textbf{UT End} & \textbf{Exp. [s]} \\ \hline
00034059002 & 13 03:59:57 & 13 17:14:02 & 2319\\ \hline
00034059003 & 14 09:59:57 & 14 20:14:26 & 1650\\ \hline
00034059004 & 15 09:53:57 & 16 16:31:29 & 5928\\ \hline
00034059005 & 18 05:49:57 & 18 14:27:33 & 1651\\ \hline
00034059006 & 19 16:40:56 & 19 19:02:54 & 2277\\ \hline
\end{tabular}
\end{table}

\section{Mass-Loss Modeling} \label{sec:variability}

\subsection{XUV Spectrum Estimation} \label{subsec:xuvEstimate}

With contemporaneous data from 2023 on WASP-69's X-ray output and WASP-69b's metastable HeI transmission, we can model WASP-69b's upper atmosphere without fully relying on proxy methods to estimate the stellar irradiation. Then, we can assess whether the magnitude of the observed metastable HeI variability is consistent with the magnitude of the stellar XUV variability. We used \texttt{p-winds} v1.4.5 \citep{dosSantos2022pwinds} for this exercise, which determines the HeI level populations with an input incident stellar spectrum. Metastable HeI is populated when the ground-state is ionized by EUV photons ($\lambda < 504$\,\AA) and subsequently recombines into the metastable state. Metastable HeI is depleted both via collisions with electrons and via ionization by mid-UV ($\lambda < 2600$\,\AA{}) photons. For giant planets orbiting K-type stars like WASP-69b, electron collisions are believed to dominate \citep{oklopcic2019radiationHe}.

We estimated WASP-69's XUV spectral energy distribution in four physically-motivated ranges:

\begin{itemize}
    \item $F_{\text{x}}$: photons from 5-41\,\AA{} observed by \textit{Swift}.
    \item $F_{\text{euv},1}$: photons from 41-504\,\AA{} that ionize the HeI singlet, ground-state H, and the metastable HeI triplet.
    \item $F_{\text{euv},2}$: photons from 504-911\,\AA{} that cannot ionize the HeI singlet but do ionize ground-state H and the metastable HeI triplet.
    \item $F_{\text{uv}}$: photons from 911-2593\,\AA{} that cannot ionize the HeI singlet or ground-state H but can ionize the metastable HeI triplet.
\end{itemize}

To estimate the total flux in the latter three bands, we determined and applied power-law scaling relations based on solar XUV measurements from \textit{TIMED/SEE} mission \citep{woods2005solarXUV}, a methodology previously applied by \cite{chadney2015UXV} and \cite{king2018xuv}. Notably, we used scaling relations based on the Sun to infer XUV bandpass fluxes for WASP-69, a K-type star. \cite{chadney2015UXV} and \cite{king2018xuv} previously considered the applicability of extrapolating these power-laws to other stellar masses, finding that the Sun's $F_{\text{euv}}/F_{\text{x}}$ versus $F_{\text{x}}$ correlation predicted values that were consistent with observations of active M-dwarfs. This evidence supports our assumption of using these power-laws to infer WASP-69's EUV flux.

For this calculation, we downloaded the Version 12 file of the mission's merged Level 3 data products that provides the daily-averaged solar flux observed by the satellite from low-Earth orbit in 10\,\AA{} bins over the range 0-1950\,\AA. Notably, reported fluxes have been slightly modified since the Version 11 release used by \cite{chadney2015UXV} and \cite{king2018xuv} due to an adjustment in the detector's estimated degradation. From these data, we filtered to only the most recent complete solar cycle: 2008 December 01 through 2019 December 01. 

\begin{table}[]
\centering
\caption{Power-law parameters and scaling factors, per the variables in Equation \ref{eq:powerLaw}, derived from solar data and used to estimate flux in XUV bandpasses for WASP-69.}
\label{tab:powerLaws}
\begin{tabular}{|c|c|c|}
\hline
Bandpass & $\alpha$ [erg\,cm$^{-2}$\,s$^{-1}$] & $\gamma$ [unitless] \\ \hline
$F_{\text{euv},1}$ & 1760 & -0.519 \\ \hline
$F_{\text{euv},2}$ & 7950 & -0.834 \\ \hline
$F_{\text{uv},1}$ & $1.70\times10^{6}$ & -0.959 \\ \hline
\end{tabular}
\end{table}

The possible wavelength coverage of the \textit{TIMED/SEE} data reaches 1950\,\AA, but values above 1900\,\AA{} are not provided for most days in our selected range. Because the data do not cover the longest wavelengths of our previously defined $F_{\text{uv}}$, we broke-up this bandpass into $F_{\text{uv},1} \in [911, 1900]$\,\AA{} and $F_{\text{uv},2} \in [1900, 2593]$\,\AA. Next, we ignored days where any bin in $F_{\text{x}}$, $F_{\text{euv},1}$, $F_{\text{euv},2}$, or $F_{\text{uv},1}$ did not have a reported flux. This strict quality cut eliminated about 1\% of days in the original range.

For each spectrum in this set of days, we summed the flux in $F_{\text{x}}$, $F_{\text{euv},1}$, $F_{\text{euv},2}$, and $F_{\text{uv},1}$. After scaling to what would be detected at the Sun's surface \citep{chadney2015UXV}, we determined power-law relationships:

\begin{equation} \label{eq:powerLaw}
    \frac{F_{\mathrm{bp}}}{F_\mathrm{x}} = \alpha (F_{\mathrm{x}})^{\gamma}\,,
\end{equation}

\noindent where $F_{\text{bp}}$ is the flux in a given UV bandpass and the best-fit $\alpha$ and $\gamma$ are constants determined for each fit. 

\begin{figure}
    \centering
    \includegraphics[width=0.95\linewidth]{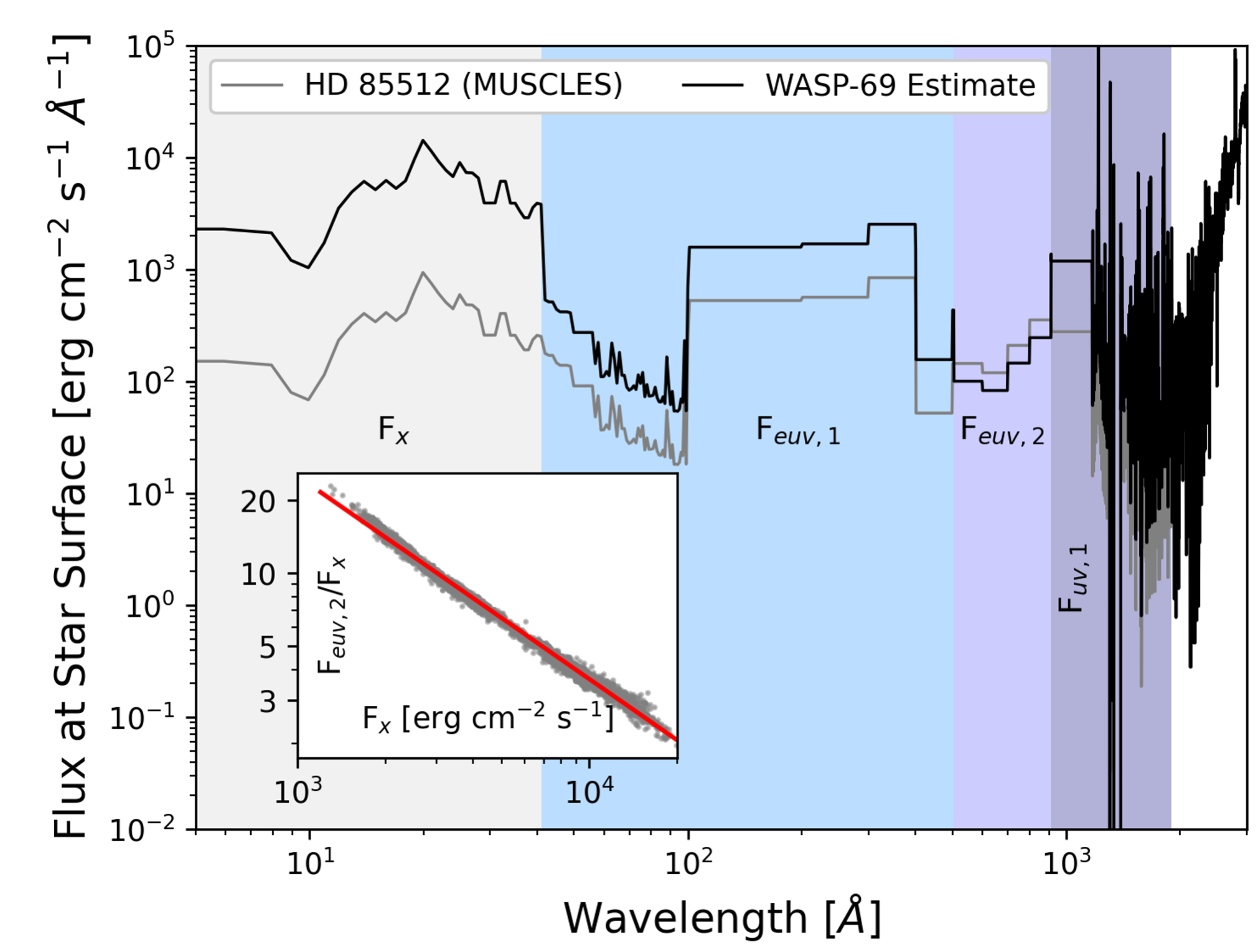}
    \caption{The main plot shows the estimated WASP-69 spectrum (black line), which was established from \textit{Swift} observations and scaling HD 85512's spectrum (grey line) to solar X-ray/UV flux ratios. Each bandpass-of-interest is shown with background shading. The inset shows the \textit{TIMED/SEE} data from Solar Cycle 24 from the $F_{\text{x}}$ and $F_{\text{euv},2}$ bands along with the best-fit power-law.}
    \label{fig:scaledHD85515}
\end{figure}

Results for each bandpass are shown in Table \ref{tab:powerLaws}. As an example, we show the best-fit power-law between $F_{\text{x}}$ and $F_{\text{euv},2}$ in the inset of Figure \ref{fig:scaledHD85515}. Importantly, the correlation between bandpassess shows that the Sun's EUV changes in concert with X-ray output across the activity cycle; we should expect X-ray variability to reliably trace EUV variability for exoplanet hosts.

To estimate the spectrum of WASP-69 in the epoch of the \textit{Swift} data and Night 2 metastable HeI data, we started from the HD 85512 spectrum published by the MUSCLES survey \citep{france2016MUSCLESI, youngblood2016MUSCLESII, loyd2016MUSCLESIII}. This star is an active K-type whose observed spectrum was first used to proxy WASP-69's XUV output by \cite{vissapragada2020He}.

We scaled the entire HD 85512 spectrum to what would be observed at the surface of the star, then scaled fluxes in each of the aforementioned ranges to be consistent with the \textit{Swift} $F_{\text{x}}$ upper bound from the X-ray observations and the power-law relationships in Table \ref{tab:powerLaws}. First, we multiplied all fluxes with wavelengths in the $F_{\text{x}}$ range such that the total flux in that band summed-up to the upper confidence bound from \textit{Swift}. For $F_{\text{euv},1}$, $F_{\text{euv},2}$, and $F_{\text{uv},1}$, we multiplied fluxes within each range by values such that the total fluxes in those bandpasses summed to the respective power-law estimates with the parameters in Table \ref{tab:powerLaws}. We left $F_{\text{uv},2}$ unchanged from the original HD 85512; the longest wavelength fluxes in TIMED/SEE barely change as $F_{\text{x}}$ varies, so we would not expect that changes in $F_{\text{x}}$ affect mid-UV fluxes.

This procedure derived an input spectrum for our atmospheric modeling routine, which we display in Figure \ref{fig:scaledHD85515}. Although spectral shapes are inconsistent between $F_{\text{x}}$ and the UV ranges, the shapes within each range are consistent with an observed star of similar mass and activity level. More importantly, this input spectrum for atmospheric modeling has fluxes in each relevant bandpass that are consistent with contemporaneous X-ray observations and scaling-relations derived from high-fidelity solar data.

\subsection{Atmospheric Variability Modeling} \label{subsec:pwinds}

Since contemporaneous X-ray data only exists for the recent epoch of metastable HeI absorption data, we constructed an atmospheric model to estimate whether the XUV change observed between 2016 and 2023 by \textit{XMM-Newton} and \textit{Swift} could explain the metastable HeI variability. We focused on comparing the GIANO Night 1 and WIRC Night 2 datasets because the magnitude of those two best-fit WIRC-like transit depths differed the most among all the metastable HeI absorption data on WASP-69b's thermosphere; if the observed stellar XUV variability could explain that observed atmospheric variability, then all metastable HeI variability could be attributed to changes across WASP-69's activity cycle. While the CARMENES Night 1 transit was collected closest in time to the \textit{XMM-Newton} visit, the \textit{XMM-Newton} observation was too time-separated from any metastable HeI data to meet our definition of contemporaneous. Thus, we used the observation with the largest variability as a basis for comparison.

We used \texttt{p-winds}, which implements the isothermal Parker wind outflow formulation by \cite{oklopcic2018windowHe} and \cite{lampon2020modelHe}. This 1D model describes an outflow for an exoplanet at the substellar point given the planetary parameters, a mass-loss rate from the substellar point $\dot{M}_{\text{sub}}$, the thermospheric temperature $T_{0}$, the H/He fraction, and an incident irradiating spectrum. To convert to a planet-averaged mass-loss rate $\dot{M}$, we divide the value inputted to \texttt{p-winds} by 4 \citep{vissapragada2022parker}.

For a reference planet-to-star radii ratio when calculating excess depths, we used the aforementioned transit depth from \textit{HST} \citep{tsiaras2018giantPlanet} that was used in Section \ref{sub:jointAll}. We assumed an atmosphere composed of 90\% H and 10\% He by number, a common modeling setup for giant planets \citep[i.e.][]{oklopcic2018windowHe,mansfield2018HATP11}, and planetary parameters from \cite{anderson2014wasp69} except for the radius $R_{p} = 1.014\,\text{R}_{\text{J}}$ from \cite{tsiaras2018giantPlanet}. In addition, we adjusted WASP-69b's gravity field in the momentum equation by the tidal effects elucidated by \cite{murrayclay2009escape} and implemented in \texttt{p-winds} by \cite{vissapragada2022parker}. Given those inputs, we found the outflow velocity and density profiles with \texttt{p-winds}.

We used the radiative transfer functionality in \texttt{p-winds} to obtain the H and He ionization fractions and level populations, assuming an incident spectrum scaled to $a/R_{\star} = 12.21$ (from Table \ref{tab:fitResults}) from the one shown on Figure \ref{fig:scaledHD85515}. After calculating the wavelength-resolved in-transit metastable HeI triplet, we Doppler shifted the model spectrum by the planet's radial velocity versus Palomar Observatory on Night 2. Finally, we convolved the simulated metastable HeI signal with the ultranarrowband filter's bandpass to determine the theoretical WIRC transit depth that would be observed for the given exoplanetary outflow parameterization. This convolution procedure is similar to the one used to create WIRC-like light curves from spectroscopic data in Section \ref{sub:jointAll}, but we are only finding the WIRC-like transit depth for one spectrum in this case.

With this model setup, we selected a fiducial value of $T_{0} = 10000$\,K and found the best-fit $\dot{M}$ via the \texttt{optimize.minimize} functionality in \texttt{scipy} with the Powell solver that matched the planet-to-star radius ratio from WIRC Night 2 from the joint fit in Section \ref{sub:jointAll}. For an energy-limited outflow, a given $(T_{0}, \dot{M})$ combination implies an efficiency $\epsilon$ via the formulation for the mass-loss rate

\begin{equation} \label{eq:massLossRate}
    \dot{M} \approx \frac{\epsilon \pi R_{p}^{3}F_{\mathrm{XUV}}}{KGM_{p}}\,,
\end{equation}

\noindent where $M_{p}$ is the planet mass, $G$ is the gravitational constant, and $K$ corrects the gravity field for the Roche lobe \citep{erakev2007rocheLobe}.

To illustrate the effect of changing XUV flux on the observed metastable HeI signal, we found the efficiency $\epsilon$ at the fiducial $T_{0}$ point with $F_{\text{XUV}}$ inferred from the \textit{Swift} upper bound $F_{\text{x}}$. Then, we held $\epsilon$ and $T_{0}$ constant while adjusting $F_{\text{x}}$ and recalculating the expected WIRC transit depths with the same \texttt{p-winds} procedure. Theoretical modeling has suggested that the constant $\epsilon$ approximation is valid for planets like WASP-69b \citep{caldrioli2021ATES1, caldrioli2022ATES2, wang2021numericalWASP69}. Notably, the terrestrial thermosphere increases in temperature with increasing solar XUV irradiation \citep{mlynczak2018TCI}, but self-consistently treating that effect is beyond the scope of $\texttt{p-winds}$. Specifically, we varied the independent variable $F_{\text{x}}$ by an order-of-magnitude from the \textit{Swift} upper bound, then estimated the EUV flux in each band from Table \ref{tab:powerLaws} for the input $F_{\text{x}}$ value according to Equation \ref{eq:powerLaw} and the procedure in Section \ref{subsec:xuvEstimate}. Next, we obtained adjusted $\dot{M}$ values from Equation \ref{eq:massLossRate} given the modified high-energy fluxes. With these new input parameters for our outflow model, we found the expected WIRC transit depth with the aforementioned \texttt{p-winds} procedure. Finally, we checked to see if these newly-constructed outflows were energetically self-consistent since excluded outflows could constrain $T_{0}$. 

\begin{figure}
    \centering
    \includegraphics[width=0.98\linewidth]{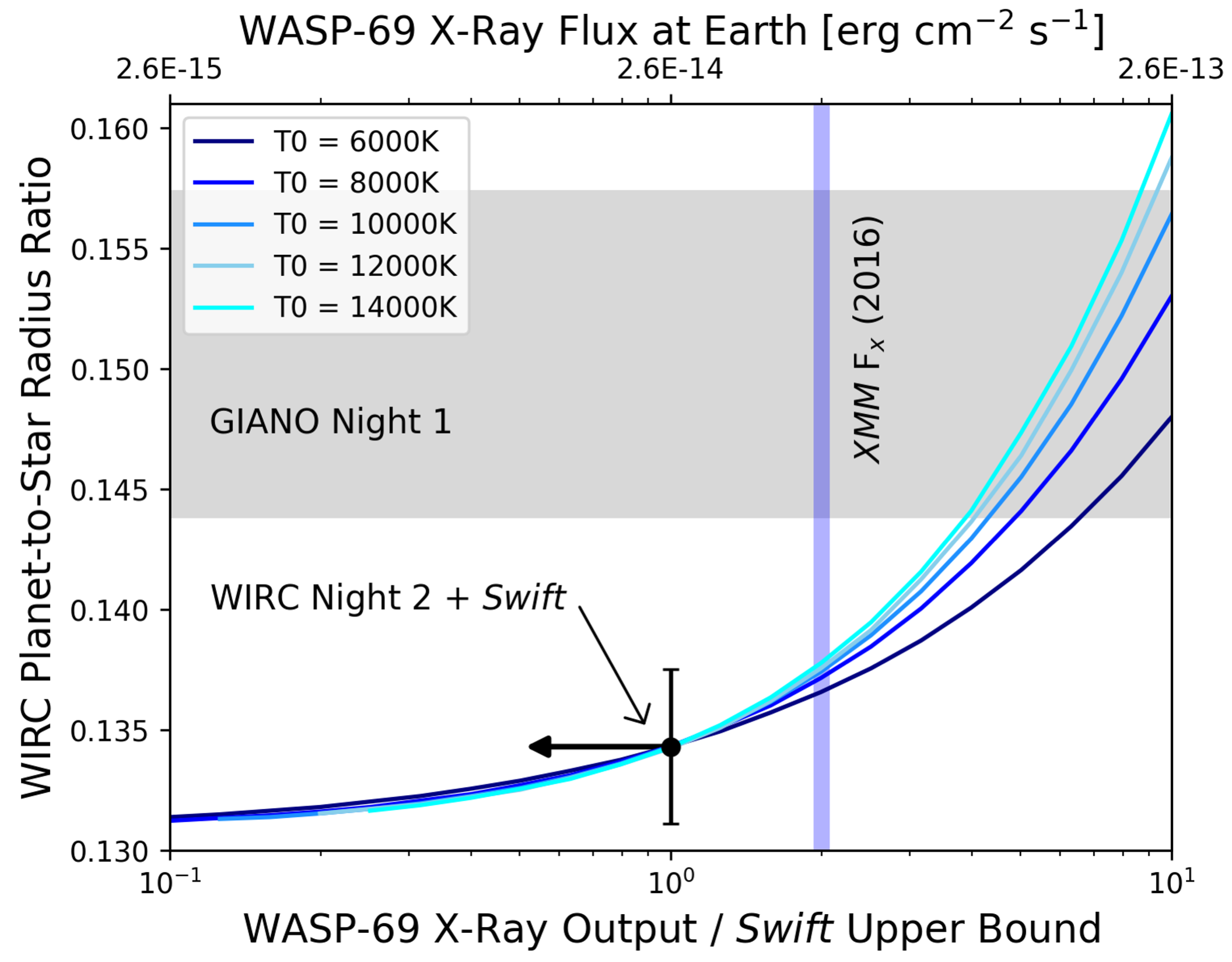}
    \caption{WIRC planet-to-star radius ratio versus WASP-69's X-ray output, normalized by the \textit{Swift} 95\% upper confidence bound. The top x-axis shows the X-ray fluxes from WASP-69 in terms of what would be measured at Earth. The black point represents the constraint from contemporaneous multi-wavelength data, where the WIRC Night 2 transit depth is taken from Section \ref{sub:jointAll}. The grey horizontal shaded bar and the blue vertical shaded bar show the WIRC-like metastable HeI transit depth from GIANO Night 1 in 2019 (best-fit value from Section \ref{sub:jointAll}) and the \textit{XMM-Newton} X-ray flux from 2016, respectively. Lines represent theoretical \texttt{p-winds} model results for various $T_{0}$ values, where the mass-loss rates are calculated with the same efficiency $\epsilon$ that is required to match the contemporaneous data from 2023. WASP-69's X-ray flux from 2016, although not contemporaneous with any metastable HeI transmission, changes in a direction and magnitude consistent with the interpretation that the drop in planetary absorption data over time stems from long-term changes in the stellar XUV across the magnetic activity cycle.}
    \label{fig:xuvChange}
\end{figure}

Figure \ref{fig:xuvChange} shows the results from this approach: the theoretical WIRC planet-to-star radius ratio as a function of X-ray flux. We ran multiple models with various assumptions on $T_{0}$ to probe the effect of changing the thermospheric temperature. Nearly all trial outflows were found to be energetically feasible, and the only inadmissible ones corresponded to low $F_{\text{x}}$ and high $T_{0}$. A version of Figure \ref{fig:xuvChange} that includes all WIRC-like metastable HeI transit depths instead of only the extreme values is located in Appendix \ref{sec:appendixModeling}.

We found that the direction of the change in metastable HeI variability -- decreasing signal for decreasing stellar X-ray flux -- is consistent with the hypothesis that the observed changes in the planetary thermosphere resulted from years-long changes in stellar XUV output. While Figure \ref{fig:xuvChange} implies that the X-ray flux during the GIANO Night 1 observations in 2019 must have been at least 2-5x larger than what \textit{XMM-Newton} observed in 2016, a critical assumption in our procedure was anchoring our \texttt{p-winds} models at the X-ray flux from \textit{Swift} 95\% confidence bound. Had we assumed a lower value for the X-ray flux on Night 2, then the intersection of the \textit{XMM-Newton} and GIANO Night 1 uncertainty ranges could have been closer to overlapping with the \texttt{p-winds} results. Importantly, the demand on WASP-69's high-energy variability implied by Figure \ref{fig:xuvChange} is consistent with broad expectations for years-long variability in stellar X-ray luminosities \citep{woods2005solarXUV}. Thus, changes in WASP-69b's metastable HeI signal are plausibly explained by changes in its host star's XUV output across the activity cycle. Future epochs with contemporaneous high-energy and metastable HeI data can add more points to Figure \ref{fig:xuvChange}, probing the correlation's strength and better testing the outflow model.

Notably, \cite{guilluy2024GAPSHeI} analyzed the velocity-resolved GIANO data from their years-long baseline (2019-2022) and found a $1.9\sigma$ significant drop in the signal from GIANO Night 1 to GIANO Night 3. That change in metastable HeI absorption is in the expected direction given a decrease in stellar XUV luminosity. While our analysis in Section \ref{sub:jointAll} found that the WIRC-like transit depths from GIANO would have appeared consistent, the nature of our convolution procedure decreases the resolution and information content of spectroscopic timeseries for the purpose of treating data consistently.

\cite{guilluy2024GAPSHeI} also presented simultaneous H$\alpha$ measurements for the 2019 and 2020 transit observations that differed. From that trend, the study suggested that the change in the metastable HeI signal between those two transits may have stemmed from WASP-69b occulting stellar regions with different amounts of activity. While that hypothesis can plausibly explain the difference between the 2019 and 2020 results, no H$\alpha$ data were taken with the 2022 transit. Thus, no similar hypothesis could be proposed to explain the additional drop in metastable HeI signal in the 2022 GIANO data. Our results, which found a tentative decrease in the metastable HeI signal from that final GIANO epoch to WIRC Night 2 in August 2023, suggest an alternative hypothesis: that the overall drop in the metastable HeI signal stems from years-long changes in stellar XUV flux across an activity cycle. In this scenario, WASP-69b's thermosphere would have changed in response to the changing high-energy radiation balance. Only by adding more epochs with contemporaneous multi-wavelength observations -- metastable HeI as well as XUV, H$\alpha$, and other stellar activity indicators -- to the dataset can these hypotheses be definitively tested.

\section{Discussion \& Conclusions} \label{sec:discussion}

Along with archival data, we have analyzed new metastable HeI absorption data from a transit of WASP-69b observed by WIRC and contemporaneous stellar X-ray data from \textit{Swift}. The recent metastable HeI transit depth differed by approximately 4$\sigma$ versus data from the same instrument four years earlier. High-resolution spectroscopy of the metastable HeI triplet in WASP-69b also shows variable absorption versus other spectroscopic data and versus the WIRC data. To interrogate the cause of this change in the planetary thermosphere, we gathered X-ray and UV data with \textit{Swift} within a few weeks of the metastable HeI observation. These data revealed that WASP-69's X-ray flux had dropped by at least a factor of two versus a snapshot in 2016, a change that is consistent with the expected magnitude of variability across stellar activity cycles and in a direction consistent with the change in stellar XUV flux required to explain the changes in metastable HeI absorption.

This study highlights the importance of two emerging frontiers in exoplanet research: multi-epoch and multi-wavelength campaigns. Looking ahead, the opportunity exists to extend the timeseries of stellar activity and planetary thermospheric state for the WASP-69 system when \textit{TESS} is scheduled to observe this target in Sector 81. Contemporaneous high-energy data from the host star and metastable HeI absorption data from the hot Jupiter in this upcoming epoch along with simultaneous broadband photometry would provide an unprecedented means to probe atmospheric outflow variability. Additional opportunity to probe outflow variability will be unlocked by forthcoming purpose-built instrumentation \citep{jentink2023night}, and pilot studies like this one can set expectations for these future campaigns. 

More broadly, our results contribute to the growing literature of metastable HeI data from exoplanet thermospheres and place WASP-69's atmosphere within the broader geoscientific discipline of aeronomy. Solar system aeronomy has flourished for decades \citep{liu2011solarIonosphereReview}, as studies of Earth \citep{mlynczak2015TCI}, Mars \citep{lee2017mavenSolarCycle}, and the gas giants \citep{jackman2011solarCycleGiantPlanet}, have traced their variable thermospheric and magnetospheric conditions to changing levels of XUV irradiation over the solar magnetic activity cycle.

In this study, we have established the feasibility of probing the same cause-and-effect process in extrasolar environments. Figure \ref{fig:aeronomy} summarizes the data that we have analyzed from the WASP-69 system and plots data from terrestrial aeronomic experiments. Solar activity is proxied as the 60-day rolling mean sunspot count\footnote{Accessed from the repository at \url{https://sidc.be/}.}. Thermospheric conditions are given by the Thermospheric Climate Index \citep[TCI;][]{mlynczak2018TCI}\footnote{Accessed from \url{https://spaceweather.com/}.}, an estimate of the total power radiated by the thermopshpere and a reliable indicator of this layer's temperature. It is apparent that the terrestrial thermosphere heats-up during solar maximum when the Sun's XUV output reaches its zenith and that the reverse occurs during solar minimum. While the WASP-69 data's cadence is sparser and the uncertainties are larger, hints of these same trends can be discerned in this star-planet system. The thermospheres of WASP-69b and Earth are in different physical regimes \citep{owen2019escapeReview}, but this contrast is instrumental to examining the role of stellar radiation and gravitational potential wells in shaping atmospheric dynamics across various environments.

A deeper understanding of exoplanet aeronomy can reveal analogous processes driven by planets' responses' to their stars, possibly answering longstanding questions on exoplanet magnetospheres \citep{oklopcic2020magneticFieldHe} and the distribution of exoplanet radii \citep{howard2010occurrence}. An important input to theories of photoevaporation is the stellar XUV irradiation \citep[e.g.][]{owen2019escapeReview}, so observing changes in planetary thermospheres as stellar XUV flux changes across the activity cycle provides a natural experiment by which to calibrate these models. Thus, studying planetary aeronomy on human timescales can help to illuminate mass-loss processes that may occur on orders-of-magnitude longer timescales.

Besides providing a means to test demographic hypotheses, exoplanets can expand the parameter space over which comparative aeronomy operates and provide critical tests of generalized atmospheric models \citep{bauer2004planetaryAeronomyBook}. Answering these questions requires combining insight from atmospheric physics and astrophysics with additional data on WASP-69 and other planetary systems. In conclusion, the nascent field of exoplanet aeronomy can illuminate these distant worlds in the context of well-studied processes from geoscience and provide hypothesis tests for widely-invoked mechanisms that may explain distributions of fundamental planetary properties.\\

\noindent{ACKNOWLEDGEMENTS}
We appreciate the detailed peer-reviewing efforts from Gloria Guilluy and a second anonymous referee -- their insightful reports helped us to improve the scientific content of this manuscript. We thank the Palomar Observatory telescope operators, support astronomers, hospitality and administrative staff, and directorate for their support. We are especially grateful to Tom Barlow, Carolyn Heffner, Diana Roderick, Kathleen Koviak, Isaac Wilson, Jennifer Milburn, and Andy Boden.

We are grateful to the \textit{Swift} team for an expedient approval of our ToO request and for executing the observations. We acknowledge the use of public data from the \textit{Swift} data archive. Parts of the results shown are based on archival observations obtained by \textit{XMM-Newton}, an ESA science mission with instruments and contributions directly funded by ESA Member States and NASA. This work has made use of the NASA Exoplanet Archive, which is operated by the California Institute of Technology, under contract with the National Aeronautics and Space Administration (NASA) under the Exoplanet Exploration Program. 

We appreciate that Romain Allart, Lisa Nortmann, Dakotah Tyler, and Gloria Guilluy shared their reduced spectra for WASP-69b from SPIRou, CAMRMENES, and Keck, respectively --- having these data allowed us to compare WIRC light curves to spectrally resolved results.

We benefited from useful conversations with Greg Laughlin, Sarbani Basu, Emma Louden, Christopher Lindsay, Earl Bellinger, Sam Cabot, Darryl Seligman, Quang Tran, Leonardo dos Santos, Fei Dai, Raissa Estrela, Jessie Christiansen, Elena Gallo, Parke Lloyd, Ethan Schreyer, Morgan Saidel, Kim Paragas, and Jonathan Gomez-Barrientos. Parts of this manuscript were checked for typographical errors and edited for clarity with assistance from OpenAI's GPT-4.

WGL gratefully acknowledges support from the Department of Defense's National Defense Science \& Engineering Graduate (NDSEG) Fellowship and Yale University's John F. Enders Fellowship for dissertation research travel. WGL also thanks the LSST-DA Data Science Fellowship Program, which is funded by LSST-DA, the Brinson Foundation, and the Moore Foundation; his participation in the program has benefited this work. ADF acknowledges funding from NASA through the NASA Hubble Fellowship grant HST-HF2-51530.001-A awarded by STScI.

\software{\texttt{numpy} \citep{harris2020numpy}, \texttt{pandas} \citep{mckinney2011pandas}, \texttt{scipy} \citep{virtanen2020scipy}, \texttt{matplotlib} \citep{hunter2007matplotlib}, \texttt{astropy} \citep{astropy:2013, astropy:2018, astropy:2022}, \texttt{photutils} \citep{bradley2023photutils}, \texttt{pymc3} \citep{pymc3}, \texttt{theano} \citep{exoplanet:theano}, \texttt{arviz} \citep{kumar2019arviz}, \texttt{exoplanet} \citep{exoplanet:joss, exoplanet:zenodo}, \texttt{ldtk} \citep{husser2013phoenix, parvianinen2015ldtk}, \texttt{starry} \citep{starry:2019}, \texttt{celerite2} \citep{exoplanet:foremanmackey17, exoplanet:foremanmackey18}, \texttt{pymc3} \citep{exoplanet:pymc3}, \texttt{corner} \citep{foremanmackey2016corner}, \texttt{p-winds} \citep{dosSantos2022pwinds}, \texttt{HEASoft} \citep{HEASoft2014}, \texttt{Xspec} \citep{Arnaud1996}.}

\bibliography{bibliography}

\begin{thebibliography}{}
\expandafter\ifx\csname natexlab\endcsname\relax\def\natexlab#1{#1}\fi
\providecommand{\url}[1]{\href{#1}{#1}}
\providecommand{\dodoi}[1]{doi:~\href{http://doi.org/#1}{\nolinkurl{#1}}}
\providecommand{\doeprint}[1]{\href{http://ascl.net/#1}{\nolinkurl{http://ascl.net/#1}}}
\providecommand{\doarXiv}[1]{\href{https://arxiv.org/abs/#1}{\nolinkurl{https://arxiv.org/abs/#1}}}

\bibitem[{{Allart} {et~al.}(2018){Allart}, {Bourrier}, {Lovis}, {Ehrenreich},
  {Spake}, {Wyttenbach}, {Pino}, {Pepe}, {Sing}, \& {Lecavelier des
  Etangs}}]{allart2018He}
{Allart}, R., {Bourrier}, V., {Lovis}, C., {et~al.} 2018, Science, 362, 1384,
  \dodoi{10.1126/science.aat5879}

\bibitem[{{Allart} {et~al.}(2019){Allart}, {Bourrier}, {Lovis}, {Ehrenreich},
  {Aceituno}, {Guijarro}, {Pepe}, {Sing}, {Spake}, \&
  {Wyttenbach}}]{allart2019WASP107}
---. 2019, \aap, 623, A58, \dodoi{10.1051/0004-6361/201834917}

\bibitem[{{Allart} {et~al.}(2023){Allart}, {Lem{\'e}e-Joliecoeur}, {Jaziri},
  {Lafreni{\`e}re}, {Artigau}, {Cook}, {Darveau-Bernier}, {Dang}, {Cadieux},
  {Boucher}, {Bourrier}, {Deibert}, {Pelletier}, {Radica}, {Benneke},
  {Carmona}, {Cloutier}, {Cowan}, {Delfosse}, {Donati}, {Doyon}, {Figueira},
  {Forveille}, {Fouqu{\'e}}, {Gaidos}, {Gu}, {H{\'e}brard}, {Kiefer},
  {K{\'o}sp{\'a}l}, {Jayawardhana}, {Martioli}, {Dos Santos}, {Shang},
  {Turner}, \& {Vidotto}}]{allart2023spirouHe}
{Allart}, R., {Lem{\'e}e-Joliecoeur}, P.~B., {Jaziri}, A.~Y., {et~al.} 2023,
  \aap, 677, A164, \dodoi{10.1051/0004-6361/202245832}

\bibitem[{{Anderson} {et~al.}(2014){Anderson}, {Collier Cameron}, {Delrez},
  {Doyle}, {Faedi}, {Fumel}, {Gillon}, {G{\'o}mez Maqueo Chew}, {Hellier},
  {Jehin}, {Lendl}, {Maxted}, {Pepe}, {Pollacco}, {Queloz}, {S{\'e}gransan},
  {Skillen}, {Smalley}, {Smith}, {Southworth}, {Triaud}, {Turner}, {Udry}, \&
  {West}}]{anderson2014wasp69}
{Anderson}, D.~R., {Collier Cameron}, A., {Delrez}, L., {et~al.} 2014, \mnras,
  445, 1114, \dodoi{10.1093/mnras/stu1737}

\bibitem[{{Arnaud}(1996)}]{Arnaud1996}
{Arnaud}, K.~A. 1996, in Astronomical Society of the Pacific Conference Series,
  Vol. 101, Astronomical Data Analysis Software and Systems V, ed. G.~H.
  {Jacoby} \& J.~{Barnes}, 17

\bibitem[{{Astropy Collaboration} {et~al.}(2013){Astropy Collaboration},
  {Robitaille}, {Tollerud}, {Greenfield}, {Droettboom}, {Bray}, {Aldcroft},
  {Davis}, {Ginsburg}, {Price-Whelan}, {Kerzendorf}, {Conley}, {Crighton},
  {Barbary}, {Muna}, {Ferguson}, {Grollier}, {Parikh}, {Nair}, {Unther},
  {Deil}, {Woillez}, {Conseil}, {Kramer}, {Turner}, {Singer}, {Fox}, {Weaver},
  {Zabalza}, {Edwards}, {Azalee Bostroem}, {Burke}, {Casey}, {Crawford},
  {Dencheva}, {Ely}, {Jenness}, {Labrie}, {Lim}, {Pierfederici}, {Pontzen},
  {Ptak}, {Refsdal}, {Servillat}, \& {Streicher}}]{astropy:2013}
{Astropy Collaboration}, {Robitaille}, T.~P., {Tollerud}, E.~J., {et~al.} 2013,
  \aap, 558, A33, \dodoi{10.1051/0004-6361/201322068}

\bibitem[{{Astropy Collaboration} {et~al.}(2018){Astropy Collaboration},
  {Price-Whelan}, {Sip{\H{o}}cz}, {G{\"u}nther}, {Lim}, {Crawford}, {Conseil},
  {Shupe}, {Craig}, {Dencheva}, {Ginsburg}, {Vand erPlas}, {Bradley},
  {P{\'e}rez-Su{\'a}rez}, {de Val-Borro}, {Aldcroft}, {Cruz}, {Robitaille},
  {Tollerud}, {Ardelean}, {Babej}, {Bach}, {Bachetti}, {Bakanov}, {Bamford},
  {Barentsen}, {Barmby}, {Baumbach}, {Berry}, {Biscani}, {Boquien}, {Bostroem},
  {Bouma}, {Brammer}, {Bray}, {Breytenbach}, {Buddelmeijer}, {Burke},
  {Calderone}, {Cano Rodr{\'\i}guez}, {Cara}, {Cardoso}, {Cheedella}, {Copin},
  {Corrales}, {Crichton}, {D'Avella}, {Deil}, {Depagne}, {Dietrich}, {Donath},
  {Droettboom}, {Earl}, {Erben}, {Fabbro}, {Ferreira}, {Finethy}, {Fox},
  {Garrison}, {Gibbons}, {Goldstein}, {Gommers}, {Greco}, {Greenfield},
  {Groener}, {Grollier}, {Hagen}, {Hirst}, {Homeier}, {Horton}, {Hosseinzadeh},
  {Hu}, {Hunkeler}, {Ivezi{\'c}}, {Jain}, {Jenness}, {Kanarek}, {Kendrew},
  {Kern}, {Kerzendorf}, {Khvalko}, {King}, {Kirkby}, {Kulkarni}, {Kumar},
  {Lee}, {Lenz}, {Littlefair}, {Ma}, {Macleod}, {Mastropietro}, {McCully},
  {Montagnac}, {Morris}, {Mueller}, {Mumford}, {Muna}, {Murphy}, {Nelson},
  {Nguyen}, {Ninan}, {N{\"o}the}, {Ogaz}, {Oh}, {Parejko}, {Parley}, {Pascual},
  {Patil}, {Patil}, {Plunkett}, {Prochaska}, {Rastogi}, {Reddy Janga},
  {Sabater}, {Sakurikar}, {Seifert}, {Sherbert}, {Sherwood-Taylor}, {Shih},
  {Sick}, {Silbiger}, {Singanamalla}, {Singer}, {Sladen}, {Sooley},
  {Sornarajah}, {Streicher}, {Teuben}, {Thomas}, {Tremblay}, {Turner},
  {Terr{\'o}n}, {van Kerkwijk}, {de la Vega}, {Watkins}, {Weaver}, {Whitmore},
  {Woillez}, {Zabalza}, \& {Astropy Contributors}}]{astropy:2018}
{Astropy Collaboration}, {Price-Whelan}, A.~M., {Sip{\H{o}}cz}, B.~M., {et~al.}
  2018, \aj, 156, 123, \dodoi{10.3847/1538-3881/aabc4f}

\bibitem[{{Astropy Collaboration} {et~al.}(2022){Astropy Collaboration},
  {Price-Whelan}, {Lim}, {Earl}, {Starkman}, {Bradley}, {Shupe}, {Patil},
  {Corrales}, {Brasseur}, {N{"o}the}, {Donath}, {Tollerud}, {Morris},
  {Ginsburg}, {Vaher}, {Weaver}, {Tocknell}, {Jamieson}, {van Kerkwijk},
  {Robitaille}, {Merry}, {Bachetti}, {G{"u}nther}, {Aldcroft},
  {Alvarado-Montes}, {Archibald}, {B{'o}di}, {Bapat}, {Barentsen}, {Baz{'a}n},
  {Biswas}, {Boquien}, {Burke}, {Cara}, {Cara}, {Conroy}, {Conseil}, {Craig},
  {Cross}, {Cruz}, {D'Eugenio}, {Dencheva}, {Devillepoix}, {Dietrich},
  {Eigenbrot}, {Erben}, {Ferreira}, {Foreman-Mackey}, {Fox}, {Freij}, {Garg},
  {Geda}, {Glattly}, {Gondhalekar}, {Gordon}, {Grant}, {Greenfield}, {Groener},
  {Guest}, {Gurovich}, {Handberg}, {Hart}, {Hatfield-Dodds}, {Homeier},
  {Hosseinzadeh}, {Jenness}, {Jones}, {Joseph}, {Kalmbach}, {Karamehmetoglu},
  {Ka{l}uszy{'n}ski}, {Kelley}, {Kern}, {Kerzendorf}, {Koch}, {Kulumani},
  {Lee}, {Ly}, {Ma}, {MacBride}, {Maljaars}, {Muna}, {Murphy}, {Norman},
  {O'Steen}, {Oman}, {Pacifici}, {Pascual}, {Pascual-Granado}, {Patil},
  {Perren}, {Pickering}, {Rastogi}, {Roulston}, {Ryan}, {Rykoff}, {Sabater},
  {Sakurikar}, {Salgado}, {Sanghi}, {Saunders}, {Savchenko}, {Schwardt},
  {Seifert-Eckert}, {Shih}, {Jain}, {Shukla}, {Sick}, {Simpson},
  {Singanamalla}, {Singer}, {Singhal}, {Sinha}, {Sip{H{o}}cz}, {Spitler},
  {Stansby}, {Streicher}, {{{S}}umak}, {Swinbank}, {Taranu}, {Tewary},
  {Tremblay}, {Val-Borro}, {Van Kooten}, {Vasovi{'c}}, {Verma}, {de Miranda
  Cardoso}, {Williams}, {Wilson}, {Winkel}, {Wood-Vasey}, {Xue}, {Yoachim},
  {Zhang}, {Zonca}, \& {Astropy Project Contributors}}]{astropy:2022}
{Astropy Collaboration}, {Price-Whelan}, A.~M., {Lim}, P.~L., {et~al.} 2022,
  \apj, 935, 167, \dodoi{10.3847/1538-4357/ac7c74}

\bibitem[{{Bauer} \& {Lammer}(2004)}]{bauer2004planetaryAeronomyBook}
{Bauer}, S.~J., \& {Lammer}, H. 2004, {Planetary aeronomy : atmosphere
  environments in planetary systems}

\bibitem[{{Becker} {et~al.}(2020){Becker}, {Gallo}, {Hodges-Kluck}, {Adams}, \&
  {Barnes}}]{becker2020TRAPPIST1}
{Becker}, J., {Gallo}, E., {Hodges-Kluck}, E., {Adams}, F.~C., \& {Barnes}, R.
  2020, \aj, 159, 275, \dodoi{10.3847/1538-3881/ab8fb0}

\bibitem[{{Behr} {et~al.}(2023){Behr}, {France}, {Brown}, {Duvvuri}, {Bean},
  {Berta-Thompson}, {Froning}, {Miguel}, {Pineda}, {Wilson}, \&
  {Youngblood}}]{behr2023muscles}
{Behr}, P.~R., {France}, K., {Brown}, A., {et~al.} 2023, \aj, 166, 35,
  \dodoi{10.3847/1538-3881/acdb70}

\bibitem[{{Bennett} {et~al.}(2023){Bennett}, {Redfield}, {Oklop{\v{c}}i{\'c}},
  {Carleo}, {Ninan}, \& {Endl}}]{bennett2023nondetectionWASP48b}
{Bennett}, K.~A., {Redfield}, S., {Oklop{\v{c}}i{\'c}}, A., {et~al.} 2023, \aj,
  165, 264, \dodoi{10.3847/1538-3881/acd34b}

\bibitem[{{Bonomo} {et~al.}(2017){Bonomo}, {Desidera}, {Benatti}, {Borsa},
  {Crespi}, {Damasso}, {Lanza}, {Sozzetti}, {Lodato}, {Marzari}, {Boccato},
  {Claudi}, {Cosentino}, {Covino}, {Gratton}, {Maggio}, {Micela}, {Molinari},
  {Pagano}, {Piotto}, {Poretti}, {Smareglia}, {Affer}, {Biazzo}, {Bignamini},
  {Esposito}, {Giacobbe}, {H{\'e}brard}, {Malavolta}, {Maldonado}, {Mancini},
  {Martinez Fiorenzano}, {Masiero}, {Nascimbeni}, {Pedani}, {Rainer}, \&
  {Scandariato}}]{bonomo2017planetMass}
{Bonomo}, A.~S., {Desidera}, S., {Benatti}, S., {et~al.} 2017, \aap, 602, A107,
  \dodoi{10.1051/0004-6361/201629882}

\bibitem[{{Bourrier} {et~al.}(2020){Bourrier}, {Wheatley}, {Lecavelier des
  Etangs}, {King}, {Louden}, {Ehrenreich}, {Fares}, {Helling}, {Llama},
  {Jardine}, \& {Vidotto}}]{bourrier2020HD189733}
{Bourrier}, V., {Wheatley}, P.~J., {Lecavelier des Etangs}, A., {et~al.} 2020,
  \mnras, 493, 559, \dodoi{10.1093/mnras/staa256}

\bibitem[{Bradley {et~al.}(2023)Bradley, Sip{\H o}cz, Robitaille, Tollerud,
  Vin{\'{\i}}cius, Deil, Barbary, Wilson, Busko, Donath, G{\"u}nther, Cara,
  Lim, Me{\ss}linger, Conseil, Bostroem, Droettboom, Bray, Bratholm, Barentsen,
  Craig, Rathi, Pascual, Perren, Georgiev, de~Val-Borro, Kerzendorf, Bach,
  Quint, \& Souchereau}]{bradley2023photutils}
Bradley, L., Sip{\H o}cz, B., Robitaille, T., {et~al.} 2023, astropy/photutils:
  1.8.0, 1.8.0,  Zenodo, \dodoi{10.5281/zenodo.7946442}

\bibitem[{{Burrows} {et~al.}(2005){Burrows}, {Hill}, {Nousek}, {Kennea},
  {Wells}, {Osborne}, {Abbey}, {Beardmore}, {Mukerjee}, {Short}, {Chincarini},
  {Campana}, {Citterio}, {Moretti}, {Pagani}, {Tagliaferri}, {Giommi},
  {Capalbi}, {Tamburelli}, {Angelini}, {Cusumano}, {Br{\"a}uninger}, {Burkert},
  \& {Hartner}}]{SwiftXRT}
{Burrows}, D.~N., {Hill}, J.~E., {Nousek}, J.~A., {et~al.} 2005, \ssr, 120,
  165, \dodoi{10.1007/s11214-005-5097-2}

\bibitem[{{Caldiroli} {et~al.}(2021){Caldiroli}, {Haardt}, {Gallo}, {Spinelli},
  {Malsky}, \& {Rauscher}}]{caldrioli2021ATES1}
{Caldiroli}, A., {Haardt}, F., {Gallo}, E., {et~al.} 2021, \aap, 655, A30,
  \dodoi{10.1051/0004-6361/202141497}

\bibitem[{{Caldiroli} {et~al.}(2022){Caldiroli}, {Haardt}, {Gallo}, {Spinelli},
  {Malsky}, \& {Rauscher}}]{caldrioli2022ATES2}
---. 2022, \aap, 663, A122, \dodoi{10.1051/0004-6361/202142763}

\bibitem[{{Chadney} {et~al.}(2015){Chadney}, {Galand}, {Unruh}, {Koskinen}, \&
  {Sanz-Forcada}}]{chadney2015UXV}
{Chadney}, J.~M., {Galand}, M., {Unruh}, Y.~C., {Koskinen}, T.~T., \&
  {Sanz-Forcada}, J. 2015, \icarus, 250, 357,
  \dodoi{10.1016/j.icarus.2014.12.012}

\bibitem[{{Corrales} {et~al.}(2021){Corrales}, {Ravi}, {King}, {May},
  {Rauscher}, \& {Reynolds}}]{corrales2021Swift}
{Corrales}, L., {Ravi}, S., {King}, G.~W., {et~al.} 2021, \aj, 162, 287,
  \dodoi{10.3847/1538-3881/ac2c67}

\bibitem[{{Czesla} {et~al.}(2022){Czesla}, {Lamp{\'o}n}, {Sanz-Forcada},
  {Garc{\'\i}a Mu{\~n}oz}, {L{\'o}pez-Puertas}, {Nortmann}, {Yan}, {Nagel},
  {Yan}, {Schmitt}, {Aceituno}, {Amado}, {Caballero}, {Casasayas-Barris},
  {Henning}, {Khalafinejad}, {Molaverdikhani}, {Montes}, {Pall{\'e}},
  {Reiners}, {Schneider}, {Ribas}, {Quirrenbach}, {Zapatero Osorio}, \&
  {Zechmeister}}]{czesla2022HATP32}
{Czesla}, S., {Lamp{\'o}n}, M., {Sanz-Forcada}, J., {et~al.} 2022, \aap, 657,
  A6, \dodoi{10.1051/0004-6361/202039919}

\bibitem[{dos Santos(2021)}]{dossantos2021review}
dos Santos, L.~A. 2021, Proceedings of the International Astronomical Union,
  17, 56–71, \dodoi{10.1017/S1743921322004239}

\bibitem[{{dos Santos} {et~al.}(2020){dos Santos}, {Ehrenreich}, {Bourrier},
  {Allart}, {King}, {Lendl}, {Lovis}, {Margheim}, {Mel{\'e}ndez}, {Seidel}, \&
  {Sousa}}]{dossantos2020WASP127}
{dos Santos}, L.~A., {Ehrenreich}, D., {Bourrier}, V., {et~al.} 2020, \aap,
  640, A29, \dodoi{10.1051/0004-6361/202038802}

\bibitem[{{dos Santos} {et~al.}(2022){dos Santos}, {Vidotto}, {Vissapragada},
  {Alam}, {Allart}, {Bourrier}, {Kirk}, {Seidel}, \&
  {Ehrenreich}}]{dosSantos2022pwinds}
{dos Santos}, L.~A., {Vidotto}, A.~A., {Vissapragada}, S., {et~al.} 2022, \aap,
  659, A62, \dodoi{10.1051/0004-6361/202142038}

\bibitem[{{Erkaev} {et~al.}(2007){Erkaev}, {Kulikov}, {Lammer}, {Selsis},
  {Langmayr}, {Jaritz}, \& {Biernat}}]{erakev2007rocheLobe}
{Erkaev}, N.~V., {Kulikov}, Y.~N., {Lammer}, H., {et~al.} 2007, \aap, 472, 329,
  \dodoi{10.1051/0004-6361:20066929}

\bibitem[{{Foreman-Mackey}(2016)}]{foremanmackey2016corner}
{Foreman-Mackey}, D. 2016, The Journal of Open Source Software, 1, 24,
  \dodoi{10.21105/joss.00024}

\bibitem[{{Foreman-Mackey}(2018)}]{exoplanet:foremanmackey18}
---. 2018, Research Notes of the American Astronomical Society, 2, 31,
  \dodoi{10.3847/2515-5172/aaaf6c}

\bibitem[{{Foreman-Mackey} {et~al.}(2017){Foreman-Mackey}, {Agol},
  {Ambikasaran}, \& {Angus}}]{exoplanet:foremanmackey17}
{Foreman-Mackey}, D., {Agol}, E., {Ambikasaran}, S., \& {Angus}, R. 2017, \aj,
  154, 220, \dodoi{10.3847/1538-3881/aa9332}

\bibitem[{{Foreman-Mackey} {et~al.}(2021){Foreman-Mackey}, {Luger}, {Agol},
  {Barclay}, {Bouma}, {Brandt}, {Czekala}, {David}, {Dong}, {Gilbert},
  {Gordon}, {Hedges}, {Hey}, {Morris}, {Price-Whelan}, \&
  {Savel}}]{exoplanet:joss}
{Foreman-Mackey}, D., {Luger}, R., {Agol}, E., {et~al.} 2021, arXiv e-prints,
  arXiv:2105.01994.
\newblock \doarXiv{2105.01994}

\bibitem[{Foreman-Mackey {et~al.}(2021)Foreman-Mackey, Savel, Luger, Agol,
  Czekala, Price-Whelan, Hedges, Gilbert, Bouma, Brandt, \&
  Barclay}]{exoplanet:zenodo}
Foreman-Mackey, D., Savel, A., Luger, R., {et~al.} 2021,
  exoplanet-dev/exoplanet v0.5.1, \dodoi{10.5281/zenodo.1998447}

\bibitem[{{France} {et~al.}(2016){France}, {Loyd}, {Youngblood}, {Brown},
  {Schneider}, {Hawley}, {Froning}, {Linsky}, {Roberge}, {Buccino},
  {Davenport}, {Fontenla}, {Kaltenegger}, {Kowalski}, {Mauas}, {Miguel},
  {Redfield}, {Rugheimer}, {Tian}, {Vieytes}, {Walkowicz}, \&
  {Weisenburger}}]{france2016MUSCLESI}
{France}, K., {Loyd}, R.~O.~P., {Youngblood}, A., {et~al.} 2016, \apj, 820, 89,
  \dodoi{10.3847/0004-637X/820/2/89}

\bibitem[{{Fulton} {et~al.}(2017){Fulton}, {Petigura}, {Howard}, {Isaacson},
  {Marcy}, {Cargile}, {Hebb}, {Weiss}, {Johnson}, {Morton}, {Sinukoff},
  {Crossfield}, \& {Hirsch}}]{fulton2017gap}
{Fulton}, B.~J., {Petigura}, E.~A., {Howard}, A.~W., {et~al.} 2017, \aj, 154,
  109, \dodoi{10.3847/1538-3881/aa80eb}

\bibitem[{{Gehrels} {et~al.}(2004){Gehrels}, {Chincarini}, {Giommi}, {Mason},
  {Nousek}, {Wells}, {White}, {Barthelmy}, {Burrows}, {Cominsky}, {Hurley},
  {Marshall}, {M{\'e}sz{\'a}ros}, {Roming}, {Angelini}, {Barbier}, {Belloni},
  {Campana}, {Caraveo}, {Chester}, {Citterio}, {Cline}, {Cropper}, {Cummings},
  {Dean}, {Feigelson}, {Fenimore}, {Frail}, {Fruchter}, {Garmire}, {Gendreau},
  {Ghisellini}, {Greiner}, {Hill}, {Hunsberger}, {Krimm}, {Kulkarni}, {Kumar},
  {Lebrun}, {Lloyd-Ronning}, {Markwardt}, {Mattson}, {Mushotzky}, {Norris},
  {Osborne}, {Paczynski}, {Palmer}, {Park}, {Parsons}, {Paul}, {Rees},
  {Reynolds}, {Rhoads}, {Sasseen}, {Schaefer}, {Short}, {Smale}, {Smith},
  {Stella}, {Tagliaferri}, {Takahashi}, {Tashiro}, {Townsley}, {Tueller},
  {Turner}, {Vietri}, {Voges}, {Ward}, {Willingale}, {Zerbi}, \&
  {Zhang}}]{SwiftMissionGRB}
{Gehrels}, N., {Chincarini}, G., {Giommi}, P., {et~al.} 2004, \apj, 611, 1005,
  \dodoi{10.1086/422091}

\bibitem[{{Guilluy} {et~al.}(2023){Guilluy}, {Bourrier}, {Jaziri}, {Dethier},
  {Mounzer}, {Giacobbe}, {Attia}, {Allart}, {Bonomo}, {Dos Santos}, {Rainer},
  \& {Sozzetti}}]{guilluy2023DREAMHe}
{Guilluy}, G., {Bourrier}, V., {Jaziri}, Y., {et~al.} 2023, \aap, 676, A130,
  \dodoi{10.1051/0004-6361/202346419}

\bibitem[{Guilluy {et~al.}(2024)Guilluy, D'Arpa, Bonomo, Spinelli, Biassoni,
  Fossati, Maggio, Giacobbe, Lanza, Sozzetti, {et~al.}}]{guilluy2024GAPSHeI}
Guilluy, G., D'Arpa, M., Bonomo, A., {et~al.} 2024, arXiv preprint
  arXiv:2403.00608

\bibitem[{{Gully-Santiago} {et~al.}(2024){Gully-Santiago}, {Morley}, {Luna},
  {MacLeod}, {Oklop{\v{c}}i{\'c}}, {Ganesh}, {Tran}, {Zhang}, {Bowler},
  {Cochran}, {Krolikowski}, {Mahadevan}, {Ninan}, {Stef{\'a}nsson},
  {Vanderburg}, {Zalesky}, \& {Zeimann}}]{gullysantiago2024variableTail}
{Gully-Santiago}, M., {Morley}, C.~V., {Luna}, J., {et~al.} 2024, \aj, 167,
  142, \dodoi{10.3847/1538-3881/ad1ee8}

\bibitem[{{Harris} {et~al.}(2020){Harris}, {Millman}, {van der Walt},
  {Gommers}, {Virtanen}, {Cournapeau}, {Wieser}, {Taylor}, {Berg}, {Smith},
  {Kern}, {Picus}, {Hoyer}, {van Kerkwijk}, {Brett}, {Haldane}, {del R{\'\i}o},
  {Wiebe}, {Peterson}, {G{\'e}rard-Marchant}, {Sheppard}, {Reddy}, {Weckesser},
  {Abbasi}, {Gohlke}, \& {Oliphant}}]{harris2020numpy}
{Harris}, C.~R., {Millman}, K.~J., {van der Walt}, S.~J., {et~al.} 2020,
  Nature, 585, 357, \dodoi{10.1038/s41586-020-2649-2}

\bibitem[{{Hoffman} \& {Gelman}(2011)}]{hoffman2011NUTS}
{Hoffman}, M.~D., \& {Gelman}, A. 2011, arXiv e-prints, arXiv:1111.4246,
  \dodoi{10.48550/arXiv.1111.4246}

\bibitem[{{Howard} {et~al.}(2010){Howard}, {Marcy}, {Johnson}, {Fischer},
  {Wright}, {Isaacson}, {Valenti}, {Anderson}, {Lin}, \&
  {Ida}}]{howard2010occurrence}
{Howard}, A.~W., {Marcy}, G.~W., {Johnson}, J.~A., {et~al.} 2010, Science, 330,
  653, \dodoi{10.1126/science.1194854}

\bibitem[{{Hunter}(2007)}]{hunter2007matplotlib}
{Hunter}, J.~D. 2007, Computing in Science and Engineering, 9, 90,
  \dodoi{10.1109/MCSE.2007.55}

\bibitem[{{Husser} {et~al.}(2013){Husser}, {Wende-von Berg}, {Dreizler},
  {Homeier}, {Reiners}, {Barman}, \& {Hauschildt}}]{husser2013phoenix}
{Husser}, T.~O., {Wende-von Berg}, S., {Dreizler}, S., {et~al.} 2013, \aap,
  553, A6, \dodoi{10.1051/0004-6361/201219058}

\bibitem[{{Jackman} \& {Arridge}(2011)}]{jackman2011solarCycleGiantPlanet}
{Jackman}, C.~M., \& {Arridge}, C.~S. 2011, \solphys, 274, 481,
  \dodoi{10.1007/s11207-011-9748-z}

\bibitem[{{Jackson} {et~al.}(2012){Jackson}, {Davis}, \&
  {Wheatley}}]{jackson2012coronalXrayAge}
{Jackson}, A.~P., {Davis}, T.~A., \& {Wheatley}, P.~J. 2012, \mnras, 422, 2024,
  \dodoi{10.1111/j.1365-2966.2012.20657.x}

\bibitem[{{Jenkins} {et~al.}(2016){Jenkins}, {Twicken}, {McCauliff},
  {Campbell}, {Sanderfer}, {Lung}, {Mansouri-Samani}, {Girouard}, {Tenenbaum},
  {Klaus}, {Smith}, {Caldwell}, {Chacon}, {Henze}, {Heiges}, {Latham},
  {Morgan}, {Swade}, {Rinehart}, \& {Vanderspek}}]{jenkins2016SPOC}
{Jenkins}, J.~M., {Twicken}, J.~D., {McCauliff}, S., {et~al.} 2016, in Society
  of Photo-Optical Instrumentation Engineers (SPIE) Conference Series, Vol.
  9913, Software and Cyberinfrastructure for Astronomy IV, ed. G.~{Chiozzi} \&
  J.~C. {Guzman}, 99133E, \dodoi{10.1117/12.2233418}

\bibitem[{{Jentink} {et~al.}(2023){Jentink}, {Bourrier}, {Lovis}, {Allart},
  {Chazelas}, {Lendl}, {Dumusque}, \& {Pepe}}]{jentink2023night}
{Jentink}, C.~F., {Bourrier}, V., {Lovis}, C., {et~al.} 2023, \mnras,
  \dodoi{10.1093/mnras/stad3285}

\bibitem[{{Johnstone} {et~al.}(2021){Johnstone}, {Bartel}, \&
  {G{\"u}del}}]{johnstone2021activeLives}
{Johnstone}, C.~P., {Bartel}, M., \& {G{\"u}del}, M. 2021, \aap, 649, A96,
  \dodoi{10.1051/0004-6361/202038407}

\bibitem[{{Kasper} {et~al.}(2020){Kasper}, {Bean}, {Oklop{\v{c}}i{\'c}},
  {Malsky}, {Kempton}, {D{\'e}sert}, {Rogers}, \&
  {Mansfield}}]{kasper2020subNeptuneHe}
{Kasper}, D., {Bean}, J.~L., {Oklop{\v{c}}i{\'c}}, A., {et~al.} 2020, \aj, 160,
  258, \dodoi{10.3847/1538-3881/abbee6}

\bibitem[{{King} {et~al.}(2018){King}, {Wheatley}, {Salz}, {Bourrier},
  {Czesla}, {Ehrenreich}, {Kirk}, {Lecavelier des Etangs}, {Louden}, {Schmitt},
  \& {Schneider}}]{king2018xuv}
{King}, G.~W., {Wheatley}, P.~J., {Salz}, M., {et~al.} 2018, \mnras, 478, 1193,
  \dodoi{10.1093/mnras/sty1110}

\bibitem[{{Kipping}(2013)}]{kipping2013limbdark}
{Kipping}, D.~M. 2013, \mnras, 435, 2152, \dodoi{10.1093/mnras/stt1435}

\bibitem[{{Kirk} {et~al.}(2022){Kirk}, {Dos Santos}, {L{\'o}pez-Morales},
  {Alam}, {Oklop{\v{c}}i{\'c}}, {MacLeod}, {Zeng}, \& {Zhou}}]{kirk2022giantHe}
{Kirk}, J., {Dos Santos}, L.~A., {L{\'o}pez-Morales}, M., {et~al.} 2022, \aj,
  164, 24, \dodoi{10.3847/1538-3881/ac722f}

\bibitem[{{Kokori} {et~al.}(2022){Kokori}, {Tsiaras}, {Edwards}, {Rocchetto},
  {Tinetti}, {Bewersdorff}, {Jongen}, {Lekkas}, {Pantelidou}, {Poultourtzidis},
  {W{\"u}nsche}, {Aggelis}, {Agnihotri}, {Arena}, {Bachschmidt}, {Bennett},
  {Benni}, {Bernacki}, {Besson}, {Betti}, {Biagini}, {Brandebourg}, {Bretton},
  {Brincat}, {Cal{\'o}}, {Campos}, {Casali}, {Ciantini}, {Crow}, {Dauchet},
  {Dawes}, {Deldem}, {Deligeorgopoulos}, {Dymock}, {Eenm{\"a}e}, {Evans},
  {Esseiva}, {Falco}, {Ferratfiat}, {Fowler}, {Futcher}, {Gaitan}, {Horta},
  {Guerra}, {Hurter}, {Jones}, {Kang}, {Kiiskinen}, {Kim}, {Laloum}, {Lee},
  {Lomoz}, {Lopresti}, {Mallonn}, {Mannucci}, {Marino}, {Mario}, {Marquette},
  {Michelet}, {Miller}, {Mollier}, {Molina}, {Montigiani}, {Mortari}, {Morvan},
  {Mugnai}, {Naponiello}, {Nastasi}, {Neito}, {Pace}, {Papadeas}, {Paschalis},
  {Pereira}, {Perroud}, {Phillips}, {Pintr}, {Pioppa}, {Popowicz}, {Raetz},
  {Regembal}, {Rickard}, {Roberts}, {Rousselot}, {Rubia}, {Savage}, {Sedita},
  {Shave-Wall}, {Sioulas}, {{\v{S}}koln{\'\i}k}, {Smith}, {St-Gelais},
  {Stouraitis}, {Strikis}, {Thurston}, {Tomacelli}, {Tomatis}, {Trevan},
  {Valeau}, {Vignes}, {Vora}, {Vra{\v{s}}{\v{t}}{\'a}k}, {Walter}, {Wenzel},
  {Wright}, \& {Z{\'\i}bar}}]{kokori2022exoClock}
{Kokori}, A., {Tsiaras}, A., {Edwards}, B., {et~al.} 2022, \apjs, 258, 40,
  \dodoi{10.3847/1538-4365/ac3a10}

\bibitem[{{Kraft} {et~al.}(1991){Kraft}, {Burrows}, \& {Nousek}}]{Kraft1991}
{Kraft}, R.~P., {Burrows}, D.~N., \& {Nousek}, J.~A. 1991, \apj, 374, 344,
  \dodoi{10.1086/170124}

\bibitem[{{Kumar} {et~al.}(2019){Kumar}, {Carroll}, {Hartikainen}, \&
  {Martin}}]{kumar2019arviz}
{Kumar}, R., {Carroll}, C., {Hartikainen}, A., \& {Martin}, O. 2019, The
  Journal of Open Source Software, 4, 1143, \dodoi{10.21105/joss.01143}

\bibitem[{{Lamp{\'o}n} {et~al.}(2020){Lamp{\'o}n}, {L{\'o}pez-Puertas}, {Lara},
  {S{\'a}nchez-L{\'o}pez}, {Salz}, {Czesla}, {Sanz-Forcada}, {Molaverdikhani},
  {Alonso-Floriano}, {Nortmann}, {Caballero}, {Bauer}, {Pall{\'e}}, {Montes},
  {Quirrenbach}, {Nagel}, {Ribas}, {Reiners}, \& {Amado}}]{lampon2020modelHe}
{Lamp{\'o}n}, M., {L{\'o}pez-Puertas}, M., {Lara}, L.~M., {et~al.} 2020, \aap,
  636, A13, \dodoi{10.1051/0004-6361/201937175}

\bibitem[{{Lee} {et~al.}(2017){Lee}, {Hara}, {Halekas}, {Thiemann},
  {Chamberlin}, {Eparvier}, {Lillis}, {Larson}, {Dunn}, {Espley}, {Gruesbeck},
  {Curry}, {Luhmann}, \& {Jakosky}}]{lee2017mavenSolarCycle}
{Lee}, C.~O., {Hara}, T., {Halekas}, J.~S., {et~al.} 2017, Journal of
  Geophysical Research (Space Physics), 122, 2768, \dodoi{10.1002/2016JA023495}

\bibitem[{{Linsky} {et~al.}(2010){Linsky}, {Yang}, {France}, {Froning},
  {Green}, {Stocke}, \& {Osterman}}]{linsky2010HD209458uv}
{Linsky}, J.~L., {Yang}, H., {France}, K., {et~al.} 2010, \apj, 717, 1291,
  \dodoi{10.1088/0004-637X/717/2/1291}

\bibitem[{{Liu} {et~al.}(2011){Liu}, {Wan}, {Chen}, \&
  {Le}}]{liu2011solarIonosphereReview}
{Liu}, L., {Wan}, W., {Chen}, Y., \& {Le}, H. 2011, Chinese Science Bulletin,
  56, 1202, \dodoi{10.1007/s11434-010-4226-9}

\bibitem[{{Loyd} {et~al.}(2016){Loyd}, {France}, {Youngblood}, {Schneider},
  {Brown}, {Hu}, {Linsky}, {Froning}, {Redfield}, {Rugheimer}, \&
  {Tian}}]{loyd2016MUSCLESIII}
{Loyd}, R.~O.~P., {France}, K., {Youngblood}, A., {et~al.} 2016, \apj, 824,
  102, \dodoi{10.3847/0004-637X/824/2/102}

\bibitem[{{Luger} {et~al.}(2019){Luger}, {Agol}, {Foreman-Mackey}, {Fleming},
  {Lustig-Yaeger}, \& {Deitrick}}]{starry:2019}
{Luger}, R., {Agol}, E., {Foreman-Mackey}, D., {et~al.} 2019, \aj, 157, 64,
  \dodoi{10.3847/1538-3881/aae8e5}

\bibitem[{{Mansfield} {et~al.}(2018){Mansfield}, {Bean}, {Oklop{\v{c}}i{\'c}},
  {Kreidberg}, {D{\'e}sert}, {Kempton}, {Line}, {Fortney}, {Henry}, {Mallonn},
  {Stevenson}, {Dragomir}, {Allart}, \& {Bourrier}}]{mansfield2018HATP11}
{Mansfield}, M., {Bean}, J.~L., {Oklop{\v{c}}i{\'c}}, A., {et~al.} 2018, \apjl,
  868, L34, \dodoi{10.3847/2041-8213/aaf166}

\bibitem[{{Mazeh} {et~al.}(2016){Mazeh}, {Holczer}, \&
  {Faigler}}]{mazeh2016neptuneDesert}
{Mazeh}, T., {Holczer}, T., \& {Faigler}, S. 2016, \aap, 589, A75,
  \dodoi{10.1051/0004-6361/201528065}

\bibitem[{McKinney {et~al.}(2011)}]{mckinney2011pandas}
McKinney, W., {et~al.} 2011, Python for high performance and scientific
  computing, 14, 1

\bibitem[{{Messina} {et~al.}(2003){Messina}, {Pizzolato}, {Guinan}, \&
  {Rodon{\`o}}}]{messina2003xray}
{Messina}, S., {Pizzolato}, N., {Guinan}, E.~F., \& {Rodon{\`o}}, M. 2003,
  \aap, 410, 671, \dodoi{10.1051/0004-6361:20031203}

\bibitem[{{Mlynczak} {et~al.}(2015){Mlynczak}, {Hunt}, {Marshall}, {Russell},
  {Mertens}, {Thompson}, \& {Gordley}}]{mlynczak2015TCI}
{Mlynczak}, M.~G., {Hunt}, L.~A., {Marshall}, B.~T., {et~al.} 2015, \grl, 42,
  3677, \dodoi{10.1002/2015GL064038}

\bibitem[{{Mlynczak} {et~al.}(2018){Mlynczak}, {Hunt}, {Russell}, \&
  {Marshall}}]{mlynczak2018TCI}
{Mlynczak}, M.~G., {Hunt}, L.~A., {Russell}, J.~M., \& {Marshall}, B.~T. 2018,
  Journal of Atmospheric and Solar-Terrestrial Physics, 174, 28,
  \dodoi{10.1016/j.jastp.2018.04.004}

\bibitem[{{Murray-Clay} {et~al.}(2009){Murray-Clay}, {Chiang}, \&
  {Murray}}]{murrayclay2009escape}
{Murray-Clay}, R.~A., {Chiang}, E.~I., \& {Murray}, N. 2009, \apj, 693, 23,
  \dodoi{10.1088/0004-637X/693/1/23}

\bibitem[{{Nasa High Energy Astrophysics Science Archive Research Center
  (Heasarc)}(2014)}]{HEASoft2014}
{Nasa High Energy Astrophysics Science Archive Research Center (Heasarc)}.
  2014, {HEAsoft: Unified Release of FTOOLS and XANADU}, Astrophysics Source
  Code Library, record ascl:1408.004.
\newblock \doeprint{1408.004}

\bibitem[{{Nortmann} {et~al.}(2018){Nortmann}, {Pall{\'e}}, {Salz},
  {Sanz-Forcada}, {Nagel}, {Alonso-Floriano}, {Czesla}, {Yan}, {Chen},
  {Snellen}, {Zechmeister}, {Schmitt}, {L{\'o}pez-Puertas}, {Casasayas-Barris},
  {Bauer}, {Amado}, {Caballero}, {Dreizler}, {Henning}, {Lamp{\'o}n}, {Montes},
  {Molaverdikhani}, {Quirrenbach}, {Reiners}, {Ribas}, {S{\'a}nchez-L{\'o}pez},
  {Schneider}, \& {Zapatero Osorio}}]{nortmann2018wasp69He}
{Nortmann}, L., {Pall{\'e}}, E., {Salz}, M., {et~al.} 2018, Science, 362, 1388,
  \dodoi{10.1126/science.aat5348}

\bibitem[{{Oklop{\v{c}}i{\'c}}(2019)}]{oklopcic2019radiationHe}
{Oklop{\v{c}}i{\'c}}, A. 2019, \apj, 881, 133, \dodoi{10.3847/1538-4357/ab2f7f}

\bibitem[{{Oklop{\v{c}}i{\'c}} \& {Hirata}(2018)}]{oklopcic2018windowHe}
{Oklop{\v{c}}i{\'c}}, A., \& {Hirata}, C.~M. 2018, \apjl, 855, L11,
  \dodoi{10.3847/2041-8213/aaada9}

\bibitem[{{Oklop{\v{c}}i{\'c}} {et~al.}(2020){Oklop{\v{c}}i{\'c}}, {Silva},
  {Montero-Camacho}, \& {Hirata}}]{oklopcic2020magneticFieldHe}
{Oklop{\v{c}}i{\'c}}, A., {Silva}, M., {Montero-Camacho}, P., \& {Hirata},
  C.~M. 2020, \apj, 890, 88, \dodoi{10.3847/1538-4357/ab67c6}

\bibitem[{{Owen}(2019)}]{owen2019escapeReview}
{Owen}, J.~E. 2019, Annual Review of Earth and Planetary Sciences, 47, 67,
  \dodoi{10.1146/annurev-earth-053018-060246}

\bibitem[{{Owen} \& {Lai}(2018)}]{owen2018subJoviandesert}
{Owen}, J.~E., \& {Lai}, D. 2018, \mnras, 479, 5012,
  \dodoi{10.1093/mnras/sty1760}

\bibitem[{{Owen} \& {Wu}(2017)}]{owen2017radiusValley}
{Owen}, J.~E., \& {Wu}, Y. 2017, \apj, 847, 29,
  \dodoi{10.3847/1538-4357/aa890a}

\bibitem[{{Paragas} {et~al.}(2021){Paragas}, {Vissapragada}, {Knutson},
  {Oklop{\v{c}}i{\'c}}, {Chachan}, {Greklek-McKeon}, {Dai}, {Tinyanont}, \&
  {Vasisht}}]{paragas2021HeHATP18}
{Paragas}, K., {Vissapragada}, S., {Knutson}, H.~A., {et~al.} 2021, \apjl, 909,
  L10, \dodoi{10.3847/2041-8213/abe706}

\bibitem[{{Parviainen} \& {Aigrain}(2015)}]{parvianinen2015ldtk}
{Parviainen}, H., \& {Aigrain}, S. 2015, \mnras, 453, 3821,
  \dodoi{10.1093/mnras/stv1857}

\bibitem[{{Pillitteri} {et~al.}(2022){Pillitteri}, {Micela}, {Maggio},
  {Sciortino}, \& {Lopez-Santiago}}]{pillitteri2022xrayHD189733}
{Pillitteri}, I., {Micela}, G., {Maggio}, A., {Sciortino}, S., \&
  {Lopez-Santiago}, J. 2022, \aap, 660, A75,
  \dodoi{10.1051/0004-6361/202142232}

\bibitem[{{Ricker} {et~al.}(2015){Ricker}, {Winn}, {Vanderspek}, {Latham},
  {Bakos}, {Bean}, {Berta-Thompson}, {Brown}, {Buchhave}, {Butler}, {Butler},
  {Chaplin}, {Charbonneau}, {Christensen-Dalsgaard}, {Clampin}, {Deming},
  {Doty}, {De Lee}, {Dressing}, {Dunham}, {Endl}, {Fressin}, {Ge}, {Henning},
  {Holman}, {Howard}, {Ida}, {Jenkins}, {Jernigan}, {Johnson}, {Kaltenegger},
  {Kawai}, {Kjeldsen}, {Laughlin}, {Levine}, {Lin}, {Lissauer}, {MacQueen},
  {Marcy}, {McCullough}, {Morton}, {Narita}, {Paegert}, {Palle}, {Pepe},
  {Pepper}, {Quirrenbach}, {Rinehart}, {Sasselov}, {Sato}, {Seager},
  {Sozzetti}, {Stassun}, {Sullivan}, {Szentgyorgyi}, {Torres}, {Udry}, \&
  {Villasenor}}]{ricker2015tess}
{Ricker}, G.~R., {Winn}, J.~N., {Vanderspek}, R., {et~al.} 2015, Journal of
  Astronomical Telescopes, Instruments, and Systems, 1, 014003,
  \dodoi{10.1117/1.JATIS.1.1.014003}

\bibitem[{{Rockcliffe} {et~al.}(2021){Rockcliffe}, {Newton}, {Youngblood},
  {Bourrier}, {Mann}, {Berta-Thompson}, {Ag{\"u}eros}, {N{\'u}{\~n}ez}, \&
  {Charbonneau}}]{rockliffe2021LyA}
{Rockcliffe}, K.~E., {Newton}, E.~R., {Youngblood}, A., {et~al.} 2021, \aj,
  162, 116, \dodoi{10.3847/1538-3881/ac126f}

\bibitem[{{Roming} {et~al.}(2005){Roming}, {Kennedy}, {Mason}, {Nousek}, {Ahr},
  {Bingham}, {Broos}, {Carter}, {Hancock}, {Huckle}, {Hunsberger}, {Kawakami},
  {Killough}, {Koch}, {McLelland}, {Smith}, {Smith}, {Soto}, {Boyd},
  {Breeveld}, {Holland}, {Ivanushkina}, {Pryzby}, {Still}, \&
  {Stock}}]{SwiftUVOT}
{Roming}, P. W.~A., {Kennedy}, T.~E., {Mason}, K.~O., {et~al.} 2005, \ssr, 120,
  95, \dodoi{10.1007/s11214-005-5095-4}

\bibitem[{Salvatier {et~al.}(2016{\natexlab{a}})Salvatier, Wiecki, \&
  Fonnesbeck}]{pymc3}
Salvatier, J., Wiecki, T.~V., \& Fonnesbeck, C. 2016{\natexlab{a}}, PeerJ
  Computer Science, 2, e55

\bibitem[{Salvatier {et~al.}(2016{\natexlab{b}})Salvatier, Wiecki, \&
  Fonnesbeck}]{exoplanet:pymc3}
---. 2016{\natexlab{b}}, PeerJ Computer Science, 2, e55

\bibitem[{{Salz} {et~al.}(2019){Salz}, {Schneider}, {Fossati}, {Czesla},
  {France}, \& {Schmitt}}]{salz2019swift}
{Salz}, M., {Schneider}, P.~C., {Fossati}, L., {et~al.} 2019, \aap, 623, A57,
  \dodoi{10.1051/0004-6361/201732419}

\bibitem[{{Salz} {et~al.}(2018){Salz}, {Czesla}, {Schneider}, {Nagel},
  {Schmitt}, {Nortmann}, {Alonso-Floriano}, {L{\'o}pez-Puertas}, {Lamp{\'o}n},
  {Bauer}, {Snellen}, {Pall{\'e}}, {Caballero}, {Yan}, {Chen}, {Sanz-Forcada},
  {Amado}, {Quirrenbach}, {Ribas}, {Reiners}, {B{\'e}jar}, {Casasayas-Barris},
  {Cort{\'e}s-Contreras}, {Dreizler}, {Guenther}, {Henning}, {Jeffers},
  {Kaminski}, {K{\"u}rster}, {Lafarga}, {Lara}, {Molaverdikhani}, {Montes},
  {Morales}, {S{\'a}nchez-L{\'o}pez}, {Seifert}, {Zapatero Osorio}, \&
  {Zechmeister}}]{salz2018HD189733He}
{Salz}, M., {Czesla}, S., {Schneider}, P.~C., {et~al.} 2018, \aap, 620, A97,
  \dodoi{10.1051/0004-6361/201833694}

\bibitem[{{Sanz-Forcada} {et~al.}(2011){Sanz-Forcada}, {Micela}, {Ribas},
  {Pollock}, {Eiroa}, {Velasco}, {Solano}, \&
  {Garc{\'\i}a-{\'A}lvarez}}]{sanzforcada2011XUV}
{Sanz-Forcada}, J., {Micela}, G., {Ribas}, I., {et~al.} 2011, \aap, 532, A6,
  \dodoi{10.1051/0004-6361/201116594}

\bibitem[{{Seager} \& {Sasselov}(2000)}]{seager2000transmission}
{Seager}, S., \& {Sasselov}, D.~D. 2000, \apj, 537, 916, \dodoi{10.1086/309088}

\bibitem[{{Shallue} \& {Vanderburg}(2018)}]{shallue2018deepLearningExoplanet}
{Shallue}, C.~J., \& {Vanderburg}, A. 2018, \aj, 155, 94,
  \dodoi{10.3847/1538-3881/aa9e09}

\bibitem[{{Spake} {et~al.}(2021){Spake}, {Oklop{\v{c}}i{\'c}}, \&
  {Hillenbrand}}]{spake2021WASP107}
{Spake}, J.~J., {Oklop{\v{c}}i{\'c}}, A., \& {Hillenbrand}, L.~A. 2021, \aj,
  162, 284, \dodoi{10.3847/1538-3881/ac178a}

\bibitem[{{Spinelli} {et~al.}(2023{\natexlab{a}}){Spinelli}, {Borsa},
  {Ghirlanda}, {Ghisellini}, \& {Haardt}}]{spinelli2023UVHabitable}
{Spinelli}, R., {Borsa}, F., {Ghirlanda}, G., {Ghisellini}, G., \& {Haardt}, F.
  2023{\natexlab{a}}, \mnras, 522, 1411, \dodoi{10.1093/mnras/stad928}

\bibitem[{{Spinelli} {et~al.}(2023{\natexlab{b}}){Spinelli}, {Gallo}, {Haardt},
  {Caldiroli}, {Biassoni}, {Borsa}, \& {Rauscher}}]{spinelli2023xray}
{Spinelli}, R., {Gallo}, E., {Haardt}, F., {et~al.} 2023{\natexlab{b}}, \aj,
  165, 200, \dodoi{10.3847/1538-3881/acc336}

\bibitem[{{Stefansson} {et~al.}(2017){Stefansson}, {Mahadevan}, {Hebb},
  {Wisniewski}, {Huehnerhoff}, {Morris}, {Halverson}, {Zhao}, {Wright},
  {O'rourke}, {Knutson}, {Hawley}, {Kanodia}, {Li}, {Hagen}, {Liu}, {Beatty},
  {Bender}, {Robertson}, {Dembicky}, {Gray}, {Ketzeback}, {McMillan}, \&
  {Rudyk}}]{stefansson2017diffuser}
{Stefansson}, G., {Mahadevan}, S., {Hebb}, L., {et~al.} 2017, \apj, 848, 9,
  \dodoi{10.3847/1538-4357/aa88aa}

\bibitem[{STScI/MAST(2021)}]{TESSFastLCs}
STScI/MAST. 2021, TESS "Fast" Light Curves - All Sectors,  STScI/MAST,
  \dodoi{10.17909/T9-ST5G-3177}

\bibitem[{{Theano Development Team}(2016)}]{exoplanet:theano}
{Theano Development Team}. 2016, arXiv e-prints, abs/1605.02688.
\newblock \url{http://arxiv.org/abs/1605.02688}

\bibitem[{{Tsiaras} {et~al.}(2018){Tsiaras}, {Waldmann}, {Zingales},
  {Rocchetto}, {Morello}, {Damiano}, {Karpouzas}, {Tinetti}, {McKemmish},
  {Tennyson}, \& {Yurchenko}}]{tsiaras2018giantPlanet}
{Tsiaras}, A., {Waldmann}, I.~P., {Zingales}, T., {et~al.} 2018, \aj, 155, 156,
  \dodoi{10.3847/1538-3881/aaaf75}

\bibitem[{{Tyler} {et~al.}(2024){Tyler}, {Petigura}, {Oklop{\v{c}}i{\'c}}, \&
  {David}}]{tyler2023wasp69}
{Tyler}, D., {Petigura}, E.~A., {Oklop{\v{c}}i{\'c}}, A., \& {David}, T.~J.
  2024, \apj, 960, 123, \dodoi{10.3847/1538-4357/ad11d0}

\bibitem[{{Vanderburg} \& {Johnson}(2014)}]{vanderburg2014K2}
{Vanderburg}, A., \& {Johnson}, J.~A. 2014, \pasp, 126, 948,
  \dodoi{10.1086/678764}

\bibitem[{{Vidal-Madjar} {et~al.}(2003){Vidal-Madjar}, {Lecavelier des Etangs},
  {D{\'e}sert}, {Ballester}, {Ferlet}, {H{\'e}brard}, \&
  {Mayor}}]{vidalmajar2003lyalpha}
{Vidal-Madjar}, A., {Lecavelier des Etangs}, A., {D{\'e}sert}, J.~M., {et~al.}
  2003, \nat, 422, 143, \dodoi{10.1038/nature01448}

\bibitem[{{Vidal-Madjar} {et~al.}(2004){Vidal-Madjar}, {D{\'e}sert},
  {Lecavelier des Etangs}, {H{\'e}brard}, {Ballester}, {Ehrenreich}, {Ferlet},
  {McConnell}, {Mayor}, \& {Parkinson}}]{vidalmajar2004metals}
{Vidal-Madjar}, A., {D{\'e}sert}, J.~M., {Lecavelier des Etangs}, A., {et~al.}
  2004, \apjl, 604, L69, \dodoi{10.1086/383347}

\bibitem[{{Virtanen} {et~al.}(2020){Virtanen}, {Gommers}, {Oliphant},
  {Haberland}, {Reddy}, {Cournapeau}, {Burovski}, {Peterson}, {Weckesser},
  {Bright}, {van der Walt}, {Brett}, {Wilson}, {Millman}, {Mayorov}, {Nelson},
  {Jones}, {Kern}, {Larson}, {Carey}, {Polat}, {Feng}, {Moore}, {VanderPlas},
  {Laxalde}, {Perktold}, {Cimrman}, {Henriksen}, {Quintero}, {Harris},
  {Archibald}, {Ribeiro}, {Pedregosa}, {van Mulbregt}, \& {SciPy 1. 0
  Contributors}}]{virtanen2020scipy}
{Virtanen}, P., {Gommers}, R., {Oliphant}, T.~E., {et~al.} 2020, Nature
  Methods, 17, 261, \dodoi{10.1038/s41592-019-0686-2}

\bibitem[{{Vissapragada} {et~al.}(2022{\natexlab{a}}){Vissapragada}, {Knutson},
  {dos Santos}, {Wang}, \& {Dai}}]{vissapragada2022parker}
{Vissapragada}, S., {Knutson}, H.~A., {dos Santos}, L.~A., {Wang}, L., \&
  {Dai}, F. 2022{\natexlab{a}}, \apj, 927, 96, \dodoi{10.3847/1538-4357/ac4e8a}

\bibitem[{{Vissapragada} {et~al.}(2020){Vissapragada}, {Knutson}, {Jovanovic},
  {Harada}, {Oklop{\v{c}}i{\'c}}, {Eriksen}, {Mawet}, {Millar-Blanchaer},
  {Tinyanont}, \& {Vasisht}}]{vissapragada2020He}
{Vissapragada}, S., {Knutson}, H.~A., {Jovanovic}, N., {et~al.} 2020, \aj, 159,
  278, \dodoi{10.3847/1538-3881/ab8e34}

\bibitem[{{Vissapragada} {et~al.}(2022{\natexlab{b}}){Vissapragada}, {Knutson},
  {Greklek-McKeon}, {Oklop{\v{c}}i{\'c}}, {Dai}, {dos Santos}, {Jovanovic},
  {Mawet}, {Millar-Blanchaer}, {Paragas}, {Spake}, {Tinyanont}, \&
  {Vasisht}}]{vissapragada2022upper}
{Vissapragada}, S., {Knutson}, H.~A., {Greklek-McKeon}, M., {et~al.}
  2022{\natexlab{b}}, \aj, 164, 234, \dodoi{10.3847/1538-3881/ac92f2}

\bibitem[{{Wagner}(1988)}]{wagner1988xrayCycle21}
{Wagner}, W.~J. 1988, Advances in Space Research, 8, 67,
  \dodoi{10.1016/0273-1177(88)90173-1}

\bibitem[{{Wang} \& {Dai}(2021)}]{wang2021numericalWASP69}
{Wang}, L., \& {Dai}, F. 2021, \apj, 914, 98, \dodoi{10.3847/1538-4357/abf1ee}

\bibitem[{{Wilson} {et~al.}(2003){Wilson}, {Eikenberry}, {Henderson},
  {Hayward}, {Carson}, {Pirger}, {Barry}, {Brandl}, {Houck}, {Fitzgerald}, \&
  {Stolberg}}]{wilson2003wirc}
{Wilson}, J.~C., {Eikenberry}, S.~S., {Henderson}, C.~P., {et~al.} 2003, in
  Society of Photo-Optical Instrumentation Engineers (SPIE) Conference Series,
  Vol. 4841, Instrument Design and Performance for Optical/Infrared
  Ground-based Telescopes, ed. M.~{Iye} \& A.~F.~M. {Moorwood}, 451--458,
  \dodoi{10.1117/12.460336}

\bibitem[{{Woods} \& {Rottman}(2005)}]{woods2005solarXUV}
{Woods}, T.~N., \& {Rottman}, G. 2005, \solphys, 230, 375,
  \dodoi{10.1007/s11207-005-2555-7}

\bibitem[{{Woods} \& {Rottman}(2002)}]{woods2002solarUVvariability}
{Woods}, T.~N., \& {Rottman}, G.~J. 2002, Geophysical Monograph Series, 130,
  221, \dodoi{10.1029/130GM14}

\bibitem[{{Wright} {et~al.}(2018){Wright}, {Newton}, {Williams}, {Drake}, \&
  {Yadav}}]{wright2018rotationActivityMdwarf}
{Wright}, N.~J., {Newton}, E.~R., {Williams}, P. K.~G., {Drake}, J.~J., \&
  {Yadav}, R.~K. 2018, \mnras, 479, 2351, \dodoi{10.1093/mnras/sty1670}

\bibitem[{{Yan} \& {Henning}(2018)}]{yan2018HalphaEscape}
{Yan}, F., \& {Henning}, T. 2018, Nature Astronomy, 2, 714,
  \dodoi{10.1038/s41550-018-0503-3}

\bibitem[{{Youngblood} {et~al.}(2016){Youngblood}, {France}, {Loyd}, {Linsky},
  {Redfield}, {Schneider}, {Wood}, {Brown}, {Froning}, {Miguel}, {Rugheimer},
  \& {Walkowicz}}]{youngblood2016MUSCLESII}
{Youngblood}, A., {France}, K., {Loyd}, R.~O.~P., {et~al.} 2016, \apj, 824,
  101, \dodoi{10.3847/0004-637X/824/2/101}

\bibitem[{{Zhang} {et~al.}(2022){Zhang}, {Cauley}, {Knutson}, {France},
  {Kreidberg}, {Oklop{\v{c}}i{\'c}}, {Redfield}, \&
  {Shkolnik}}]{zhang2022hd189733}
{Zhang}, M., {Cauley}, P.~W., {Knutson}, H.~A., {et~al.} 2022, \aj, 164, 237,
  \dodoi{10.3847/1538-3881/ac9675}

\bibitem[{{Zhang} {et~al.}(2023){Zhang}, {Knutson}, {Dai}, {Wang}, {Ricker},
  {Schwarz}, {Mann}, \& {Collins}}]{zhang2023miniNeptunes}
{Zhang}, M., {Knutson}, H.~A., {Dai}, F., {et~al.} 2023, \aj, 165, 62,
  \dodoi{10.3847/1538-3881/aca75b}

\end{thebibliography}
\bibliographystyle{aasjournal}

\appendix

\section{Robustness of the Transit Depth when Fitting to WIRC Nights Individually} \label{sec:appendixRobustness}

In this Appendix, we further interrogate our individual light curve models for each night of Palomar/WIRC transit photometry with the metastable HeI filter. The purpose of these tests is to assess whether time-correlated noise plagues the fits and to examine the transit depth variability's robustness with respect to methods in the fitting procedure, such as the detrending covariates, aperture size, and length of out-of-transit baseline. All \texttt{NUTS} implementations in this Appendix use the same hyperparameters as the ones in Section \ref{subsec:individual}. For both nights, we found that the effect of detrending on centroid distance was small versus the effects of detrending on either the water proxy or airmass. Therefore, we ignored this covariate in these analyses.

\subsection{Light Curves of Individually-Fit Transits from Palomar/WIRC}

\begin{figure*}[!h]
    \centering
    \includegraphics[width=0.9\linewidth]{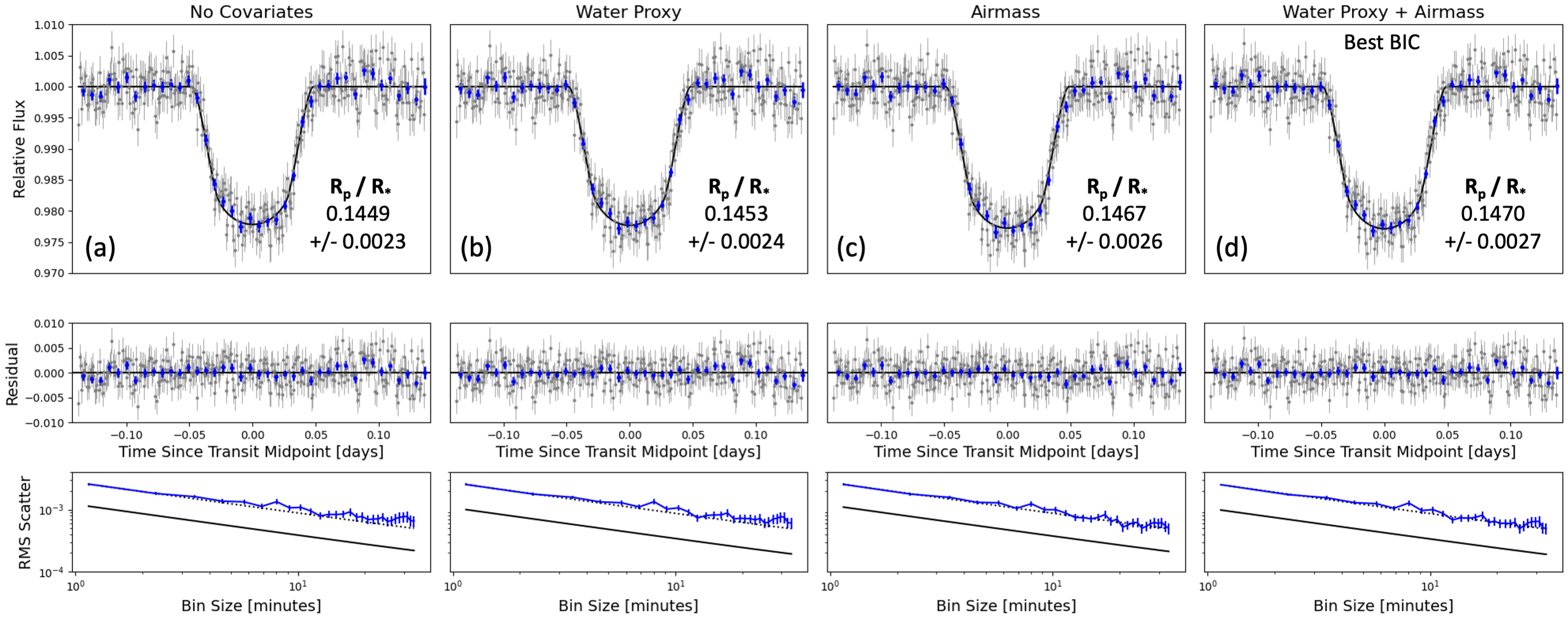}
    \caption{The top panels show the best-fit light curves for Night 1 Palomar/WIRC photometry with different sets of detrending covariates. The fit with the best BIC from Section \ref{subsec:individual} is marked at the top of its light curve. Each light curve is annotated with the derived $(R_{p}/R_{\star})$ from the given model. The middle panels show the residuals as the best-fit model subtracted from the detrended photometry. Background (grey) points on both panels derive from the individual images, and blue points are binned to ten-minute intervals. The bottom panels show the Allan deviation plots for each light curve model. Colored lines with errorbars denote the rms error of the binned residuals for each light curve. Solid black lines denote the expected noise from purely Poisson statistics, and dotted lines shows the photon noise scaled-up to the rms error of the binned residuals.}
    \label{fig:night1Lightcurves}
\end{figure*}

\begin{figure*}[!h]
    \centering
    \includegraphics[width=0.9\linewidth]{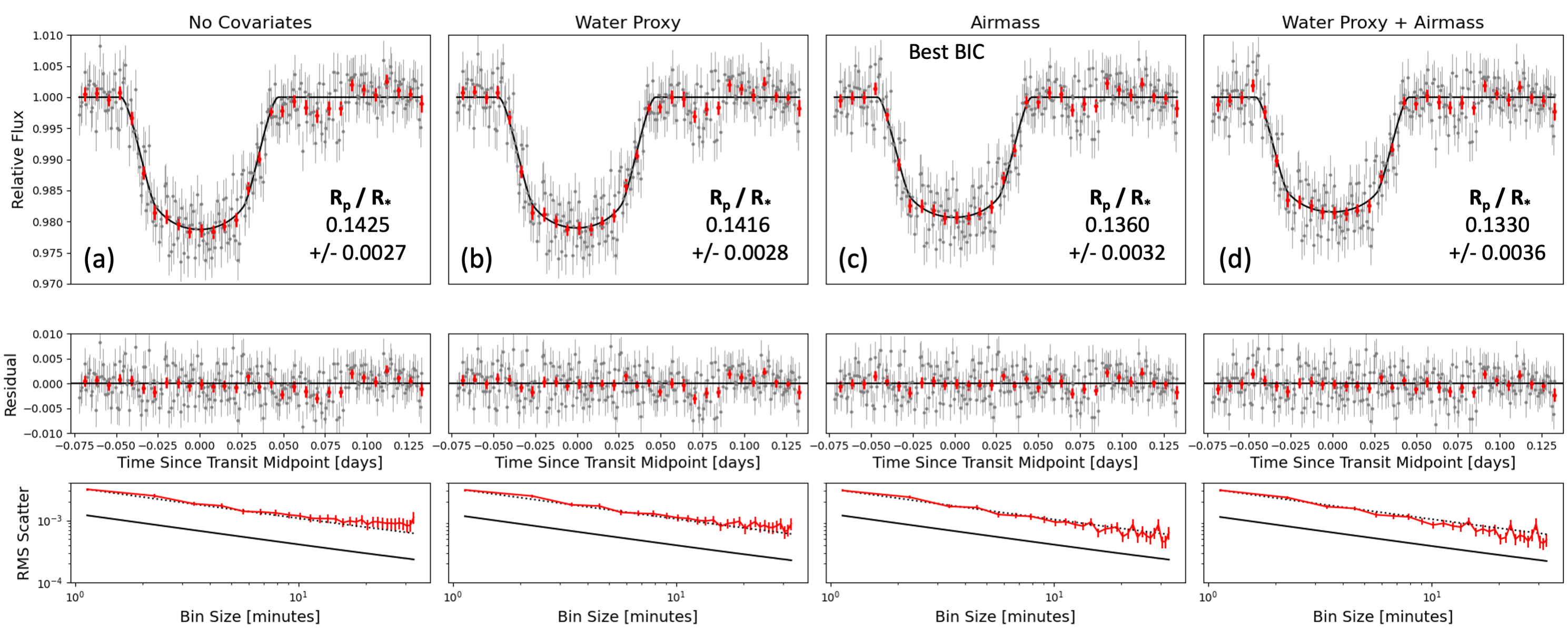}
    \caption{Analogous plots to Figure \ref{fig:night1Lightcurves} for the Night 2 WIRC photometry.}
    \label{fig:night2Lightcurves}
\end{figure*}

Here, we display the best-fit light curves, timeseries of residuals, and Allan deviation plots from fitting each night of WIRC photometry individually. Figures \ref{fig:night1Lightcurves} and \ref{fig:night2Lightcurves} show light curves from Nights 1 and 2, respectively. For each timeseries, we used the optimal aperture as selected in Section \ref{subsec:individual} --- 11 and 12 pixels for Nights 1 and 2, respectively --- but permitted free limb-darkening coefficients in each fit. We denote the fit with the best BIC (see Figure \ref{fig:allBICs} for a plot of normalized BIC values) for which we reported the derived $(R_{p}/R_{\star})_{\text{W}}$ in Section \ref{subsec:individual}.

The Night 1 transit depths are consistent within 1$\sigma$ for all relevant sets of covariates. Moreover, the derived weights on the water proxy and airmass are small enough to not visibly alter the structure in the residuals. Detrending on airmass removes some degree of correlated noise, but none of the Allan deviation plots exceed the expected noise levels scaled-up by the derived timeseries jitter beyond the uncertainties.

From Figure \ref{fig:night2Lightcurves}, it is evident that the Night 2 light curve should be detrended on airmass to control time-correlated noise. With time as the only covariate, the post-egress photometry consists of six consecutive binned points below the best-fit line followed by six consecutive binned points above the best-fit line. This fit was not favored by the BIC anyways, but the BIC also does not penalize time-correlated noise. Thus, this model is likely more disfavored than suggested by our BIC comparison methodology. The airmass curve and WASP-69 light curve look qualitatively similar, reaching their minima near the same time, so detrending on airmass alters the derived transit depth.

\subsection{Dependence of Recovered Transit Depth on Aperture Size and Covariate Set}

\begin{figure*}[!h]
    \centering
    \includegraphics[width=0.8\linewidth]{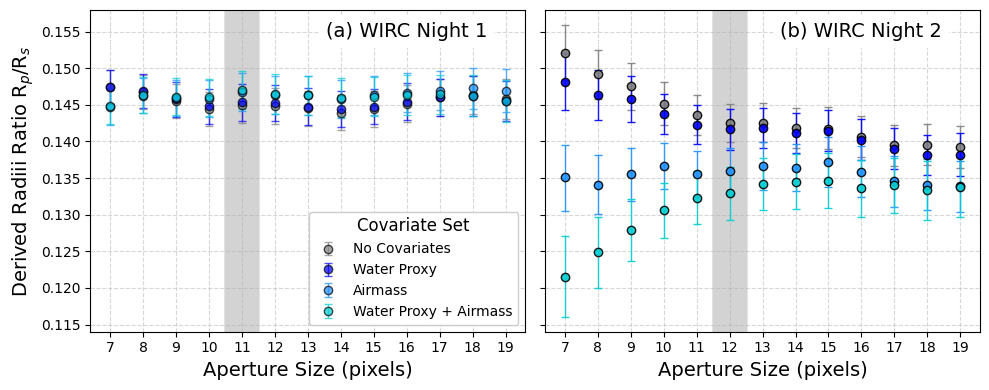}
    \caption{Derived planet-to-star radius ratios $(R_{p}/R_{\star})$ versus aperture size from fitting WIRC photometry with various sets of detrending covariates. Error bars indicate 1$\sigma$ uncertainties. Vertical shading highlights the aperture size that minimized the rms scatter as described in Section \ref{subsec:individual}. In addition to the covariates listed in the legend, all light curves are detrended by time.}
    \label{fig:changeCovariates}
\end{figure*}

\begin{figure*}[!h]
    \centering
    \includegraphics[width=0.8\linewidth]{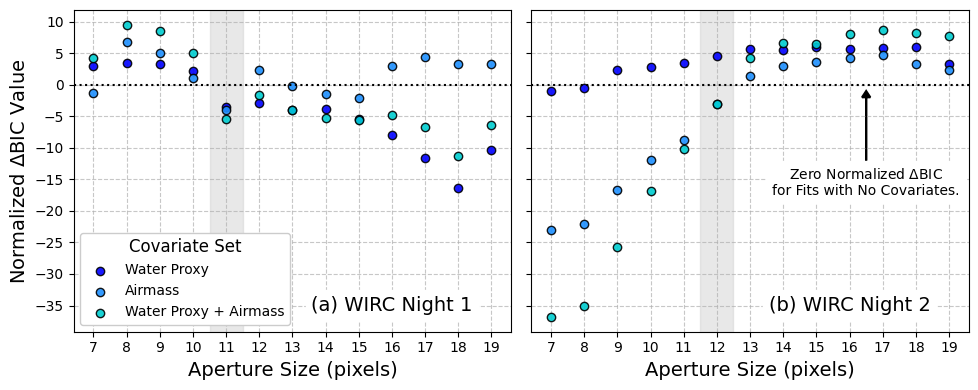}
    \caption{Normalized $\Delta{\text{BIC}}$ versus aperture size for candidate sets of covariates. For each aperture size, the $\Delta\text{BIC}$ values are computed relative to the BIC of the best-fit model without any covariates other than a linear trend with time. The most negative values of $\Delta\text{BIC}$ for each aperture size indicate the most preferred set of detrending covariates by that model selection criterion, but can only be compared to the other $\Delta\text{BIC}$ values at the same aperture size. Vertical shading highlights the aperture size that minimized the rms scatter, as described in Section \ref{subsec:individual}.}
    \label{fig:allBICs}
\end{figure*}

Here, we examine light curve modeling results for a range of aperture sizes and combinations of possibly relevant covariates. Figure \ref{fig:changeCovariates} plots the best-fit planet-to-star radii ratios at various aperture sizes, allowing free limb-darkening coefficients for each model. To assist with the interpretation, Figure \ref{fig:allBICs} shows the BIC for each fit relative to the BIC at each aperture size where no covariates other than time are considered ($\Delta\text{BIC}$). To facilitate the BIC comparison, fixed limb-darkening coefficients were used in Figure \ref{fig:allBICs} (see Section \ref{subsec:individual}). Together, these plots demonstrate changes in the best-fit transit depth and model preference and over a broad range of methodological parameter space.

Night 1 results are insensitive to aperture size and the covariates on which the light curve is detrended; all $(R_{p}/R_{\star})$ values on Figure \ref{fig:changeCovariates} are consistent within their 1$\sigma$ bounds. The globally preferred light curve from the methodology of Section \ref{subsec:individual} was detrended on the water proxy and airmass with an 11 pixel aperture. In contrast, the derived Night 2 transit depths do change for different covariates. This effect is especially apparent for small aperture sizes, but all those light curves had large rms scatter in the residuals and were not globally preferred. Even though covariate choice is less important for the transit depth determination when using large apertures, those models also result in large rms scatter in the residuals. For 1-2 pixel changes in aperture size, keeping a constant covariate set does not change $(R_{p}/R_{*})$ beyond the 1$\sigma$ uncertainties. At the optimized Night 2 aperture size of 12 pixels, either detrending on airmass alone or the water proxy and airmass are nearly equally preferred in $\Delta\text{BIC}$ terms. The preferred fits for larger apertures on Night 2 are the light curves without any covariates except for a linear trend in time.

\subsection{Effect of Time Baseline on Night 1 Light Curve}

A notable difference between the observations on Night 1 and Night 2 was the latter timeseries' comparatively short pre-ingress baseline. While the Night 1 timeseries consists of 345 total science exposures, the Night 2 timeseries contains 83 fewer points (262 images). The post-egress baselines are nearly the same length, so the difference in phase coverage is almost entirely in the pre-ingress baseline. Importantly, WASP-69b's transit itself was fully captured in the observations for both nights.

To examine whether the shorter pre-ingress baseline may affect the reliability of the Night 2 planet-to-star radius ratio determination, we checked whether $(R_{p}/R_{\star})$ derived from the Night 1 photometry is affected by excluding data from any of the pre-ingress, in-transit, and post-egress timeseries. We re-ran the fitting methodology described in Section \ref{subsec:individual} for the Night 1 data, but removed 83 points from the timeseries so that this modified Night 1 timeseries matched the length of the Night 2 data. We started by running a light curve fit in which we removed the first 83 points to cut the pre-ingress baseline (``Shortened Pre-Ingress Baseline" in Table \ref{tab:changeBaseline}). In our second fit, we instead removed the 83 points that surrounded the transit midpoint (``Excluded Mid-Transit Points" in Table \ref{tab:changeBaseline}). For the final fit, we instead removed the final 83 points to cut part of the post-ingress baseline (``Shortened Post-Egress Baseline" in Table \ref{tab:changeBaseline}). We kept constant limb-darkening coefficents for the final fits $[u_{1}, u_{2}] = [0.38, 0.12]$ for consistency between trials.

The resulting $(R_{p}/R_{\star})$ values are provided in Table \ref{tab:changeBaseline}. Removing the out-of-transit baseline did not affect $(R_{p}/R_{\star})$ by more than the 1$\sigma$ uncertainties. Removing points from the middle of the transit gave a transit depth in tension with the full timeseries value, but only at $1.2\sigma$. While the exercise of shortening the Night 1 timeseries could not exclude the possibility of a systematic transit depth offset in Night 2, the results suggest that the Night 1 data has sufficient out-of-transit baseline for a good depth determination.

\begin{table*}[!h]
\centering
\caption{Results of fitting Night 1 photometry with shortened timeseries of the same length as the full Night 2 timeseries.}
\label{tab:changeBaseline}
\begin{tabular}{|c|c|c|c|}
\hline
\textbf{Description} & \textbf{Number of Points} & \textbf{Aperture} & \textbf{Best-Fit $(R_{p}/R_{\star})$} \\ \hline
Full, Unmodified Timeseries & 345 & 11 pixels & 0.1474 $\pm$ 0.0017 \\ \hline
Shortened Pre-Ingress Baseline & 262 & 10 pixels & 0.1471 $\pm$ 0.0018\\ \hline
Excluded Mid-Transit Points & 262 & 9 pixels & 0.1437 $\pm$ 0.0025\\ \hline
Shortened Post-Egress Baseline & 262 & 11 pixels & 0.1466 $\pm$ 0.0017\\ \hline
\end{tabular}
\end{table*}

\section{Figures from Joint Fitting All Metastable He I Data} \label{sec:appendixCorner}

\begin{figure*}[!h]
    \centering \includegraphics[width=0.83\linewidth]{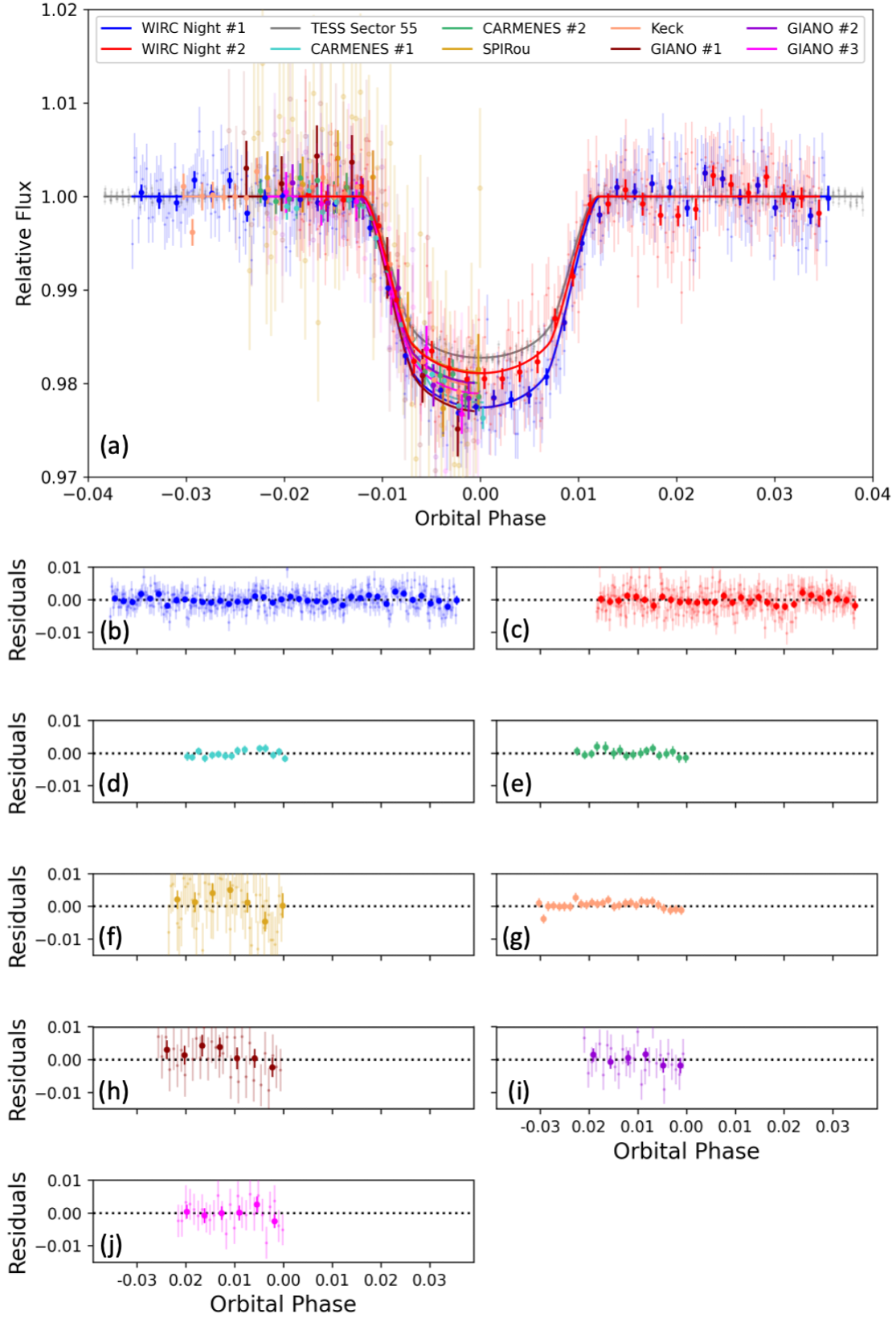}
    \caption{Panel (a) shows the best-fit, phase-folded light curve models from the procedure in Section \ref{sub:jointAll}. Panels (b), (c), (d), (e), (f), (g), (h), (i), and (j) show the residuals of the metastable HeI light curve models versus the best-fit light curve from panel (a). These residual plots correspond to WIRC Night 1, WIRC Night 2, CARMENES Night 1, CARMENES Night 2, SPIRou, Keck, GIANO Night 1, GIANO Night 2, and GIANO Night 3, respectively; the legend in panel (a) also corresponds to the color-coding in panels (b)-(j). This figure is the analog of Figure \ref{fig:jointFit} from Section \ref{subsec:jointTESS} for the joint fit from Section \ref{sub:jointAll}. }
    \label{fig:jointFitAll}
\end{figure*}

\begin{figure*}[!h]
    \centering \includegraphics[width=1\linewidth]{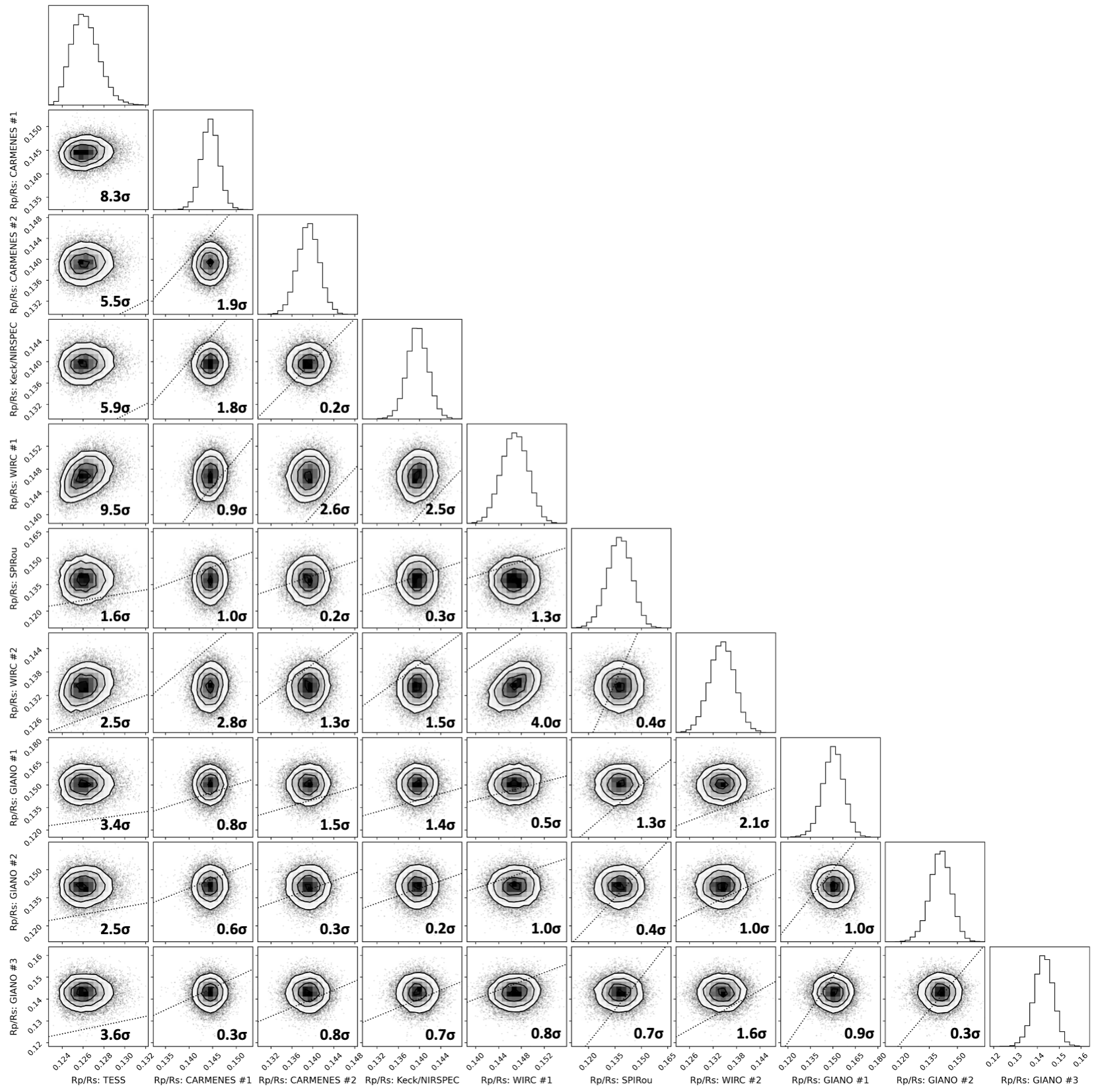}
    \caption{Corner plot of the planet-to-star radii ratios for all light curves from the joint fit in Section \ref{sub:jointAll} that included broadband \textit{TESS} observations with metastable HeI data from both ultra-narrowband photometry and spectroscopy. Each of the 2D histograms is annotated with the difference (in $\sigma$) between the two parameters as derived from the posterior samples themselves. Dotted black lines on each 2D histogram indicates where the two radii are equal.}
    \label{fig:cornerSpectralRadii}
\end{figure*}

Here, we display the best-fit light curve models and flux residuals (Figure \ref{fig:jointFitAll}) as well as a corner plot (Figure \ref{fig:cornerSpectralRadii}) of posterior samples for all planet-to-star radii ratios from the joint fitting procedure in Section \ref{sub:jointAll}. In Figure \ref{fig:cornerSpectralRadii}, we annotate each panel with the sigma difference between the two $(R_{p}/R_{*})$ values from the \texttt{NUTS} runs and denote the line where the two values would be equal. All WIRC-like metastable HeI light curves except the SPIRou data (with large uncertainties) are at least $2\sigma$ discrepant from the \textit{TESS} broadband light curve. 

It is apparent that the second WIRC transit, taken nearly four years after any other metastable HeI data, is shallower than all other metastable HeI light curves. The best-fit SPIRou $(R_{p}/R_{*})$ is closest to the result for WIRC Night 2, but the uncertainty on SPIRou is large enough that this result is effectively uninformative. Aside from the two WIRC transit depths, the posterior distributions for most of the $(R_{p}/R_{*})$ pairs are not strongly correlated. This result indicates that changes in the best-fit ephemeris and shape affect WIRC transit depths more than other timeseries.

\section{Additional Figures from \texttt{p-winds} Modeling} \label{sec:appendixModeling}

Here, we show another version of Figure \ref{fig:xuvChange} with all the metastable HeI data that was used in Section \ref{sub:jointAll} for comparison to the results of the \texttt{p-winds} modeling from Section \ref{subsec:pwinds}. Only the WIRC Night 2 observations in August 2023 were accompanied by contemporaneous X-ray measurements of WASP-69's high-energy luminosity; this epoch is the only one that can be represented by a point on this plot.

\begin{figure*}[!h]
    \centering \includegraphics[width=0.8\linewidth]{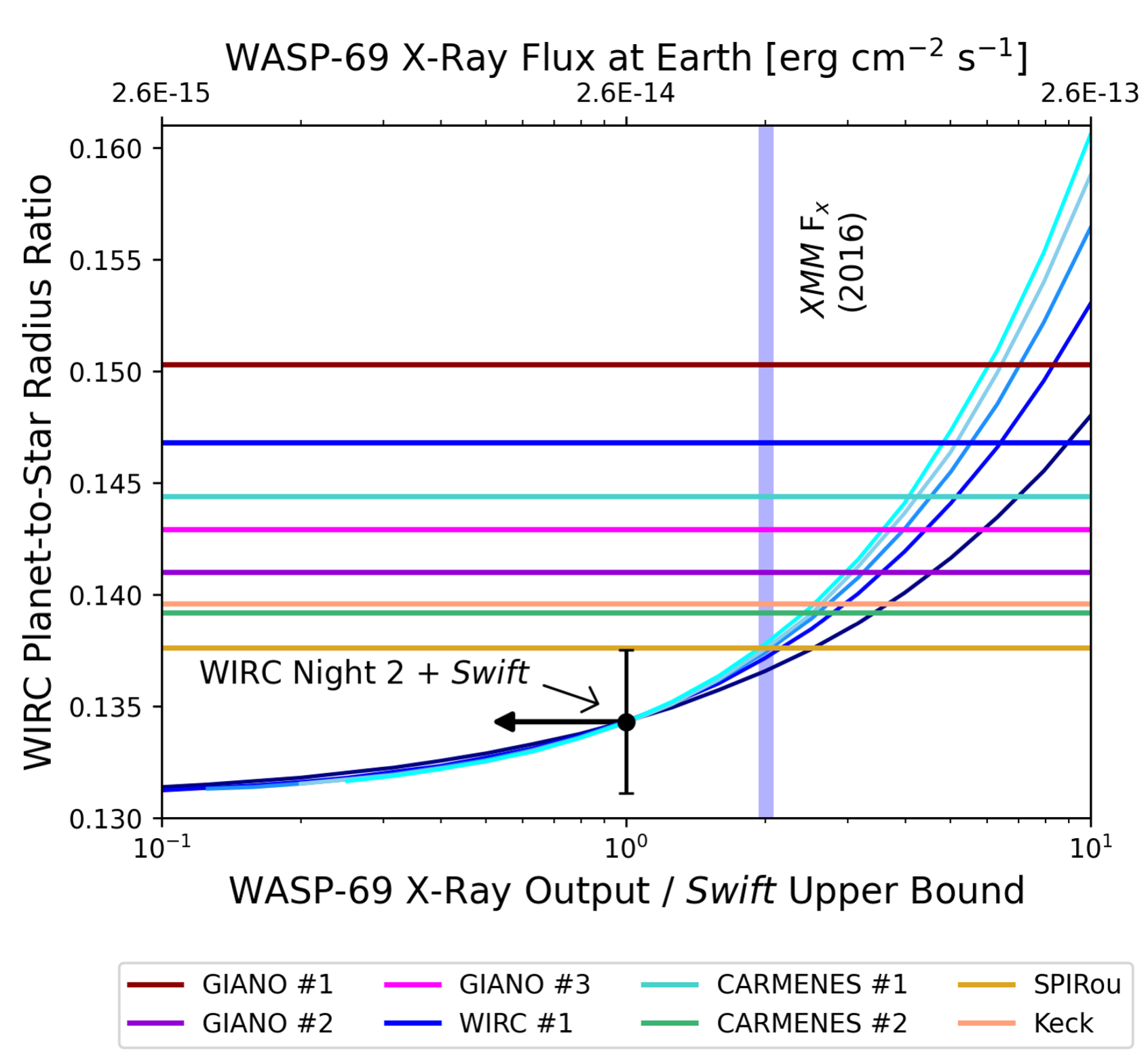}
    \caption{Analogous plot to Figure \ref{fig:xuvChange} that includes all the WIRC-like metastable HeI transits instead of only the extreme values. For each WIRC-like metastable HeI lightcurve without contemporaneous XUV data, we display a horizontal line at the best-fit planet-to-star radii ratio from the joint fit in Section \ref{sub:jointAll}. Uncertainty bounds on the WIRC-like metastable HeI transit depths are not displayed on the plot but are presented in Table \ref{tab:transitDepthsAll}.}
    \label{fig:xuvChangeAll}
\end{figure*}

\end{document}